\documentclass[fleqn,10pt]{wlscirep}
\usepackage[utf8]{inputenc}
\usepackage[T1]{fontenc}
\title{Enhancing Predictive Accuracy in Tennis: Integrating Fuzzy Logic and CV-GRNN for Dynamic Match Outcome and Player Momentum Analysis}
\usepackage{amsmath}  
\usepackage{graphicx} 
 \usepackage{multirow}
  \usepackage{setspace}
 \usepackage{listings}
 \usepackage{xcolor}
\usepackage{pythonhighlight}
 \usepackage{colortbl}
 \usepackage{tabularray}
 \usepackage{hyperref} 
 \usepackage{color}
 \usepackage{float}
 \usepackage{amssymb}
  \usepackage{setspace}
 \usepackage{color}
 \usepackage{float}
 \usepackage{amssymb}
 \usepackage{amsmath}
 \usepackage{subcaption}
 \usepackage{algorithm}
\usepackage{algpseudocode}
\usepackage{arydshln}
\usepackage{lipsum} 
\usepackage{graphicx}
\usepackage{eso-pic}
\usepackage{enumitem}

\usepackage[numbers,sort&compress]{natbib} 
\author[1,*]{Kechen Li}
\author[2,*]{Jiaming Liu}
\author[3]{Zhenyu Wu}
\author[1,$\dagger$]{Tianbo Ji}
\affil[1]{School of Transportation and Civil Engineering, Nantong University, China, 226000}
\affil[2]{Mathematics college, Nanjing University of Aeronautics and Astronautics, Nanjing, China, 211100}
\affil[3]{College of Aerospace Engineering, Nanjing University of Aeronautics and Astronautics, Nanjing, China, 211100}
\affil[*]{Equal contribution.}
\affil[$\dagger$]{Corresponding author: jitianbo@ntu.edu.cn}

\begin{abstract}
The predictive analysis of match outcomes and player momentum in professional tennis has long been a subject of scholarly debate. In this paper, we introduce a novel approach to game prediction by combining a multi-level fuzzy evaluation model with a CV-GRNN model. We first identify critical statistical indicators via Principal Component Analysis and then develop a two-tier fuzzy model based on the Wimbledon data.
In addition, the results of Pearson Correlation Coefficient indicate that the momentum indicators, such as Player Win Streak and Score Difference, have a strong correlation among them, revealing insightful trends among players transitioning between losing and winning streaks. 
Subsequently, we refine the CV-GRNN model by incorporating 15 statistically significant indicators, resulting in an increase in accuracy to 86.64\% and a decrease in MSE by 49.21\%. This consequently strengthens the methodological framework for predicting tennis match outcomes, emphasizing its practical utility and potential for adaptation in various athletic contexts.

\end{abstract}

\begin{document}

\flushbottom
\maketitle
%
%
\thispagestyle{empty}

\section{INTRODUCTION}
Tennis is a sport renowned for its complex and dynamic nature, influenced by a multitude of factors that collectively determine the outcome of a match. These factors range from the physical prowess and technical skills of individual players to their strategic acumen and psychological resilience. 
Traditionally, the analysis of tennis performance relies on statistical methods, including grey correlation, non-balance compensation, game theory, and big data mining \cite{a01,a02,a03,a04}. 
Recently, machine learning and deep learning techniques enable the analysis of complex patterns and relationships within vast datasets, and it can improve predictive analytics by taking into account player fatigue, historical performance, and real-time match conditions \cite{a05,a06,a07,a08,a09,a11,a12,a13}.

However, existing deep learning models generally focused on individual player metrics \cite{a14,a15}, which overlooks the critical interplay between competitors -- which is often referred to as ``momentum'' -- during a match. Such oversight is problematic momentum can dramatically influence the trajectory of a match. Momentum, which is characterized by streaks, consecutive scores, and score differences, is a crucial yet under-explored aspect of tennis analytics. 

In this paper, we aim to investigate the approach to the development of a robust model for predicting player performance with emphasis on capturing player momentum \cite{a16,a17,a18,a19,a20}. Therefore, we propose a novel hybrid evaluation model which integrates a multi-level fuzzy comprehensive evaluation framework with an optimized Generalized Regression Neural Network (GRNN). By additionally leveraging the cross-validation (CV) techniques, our model -- CV-GRNN -- is capable of mitigating over-fitting and enhancing predictive robustness. Our model is capable of considering both individual metrics and momentum, enabling the systematic assessment of player performances. And our model incorporates momentum-based metrics to capture the dynamic nature of tennis matches, thereby improving prediction accuracy.

The rest of the paper is organized as follows.   Section \ref{sec:RELATED WORK} surveys the landscape of existing methodologies for predicting tennis match outcomes, highlighting the transition from conventional statistical models to contemporary machine learning approaches.   Section \ref{sec:Method}  introduces our proposed model that integrates fuzzy logic with a CV-GRNN, outlining the theoretical framework, the motivation for selecting critical performance metrics, and providing insights into the Wimbledon dataset's role in our analysis.   Section \ref{EXPERIMENTS AND ANALYSES} delineates the data collection methodology and presents experimental results, illustrating the superior predictive accuracy of our model, which achieves an accuracy rate of 86.64\% and reduces the mean square error by 49.21\% compared to existing models, along with an in-depth analysis of the model's predictive capabilities.   Section \ref{Discussion}  reflects on the significance of our findings, contextualizing our approach among prior research, and contemplating both current limitations and prospective avenues for future inquiry.   The paper culminates in Section \ref{Conclusions}, which offers a synthesis of our key contributions and underscores the pragmatic value of our predictive model in the realm of tennis analytics.

\section{RELATED WORK}
\label{sec:RELATED WORK}

Evaluating player performance\cite{a41}  in tennis has been extensively researched, with various methodologies used to analyze and predict game outcomes\cite{a42}. Traditional approaches have often relied on statistical methods, including the grey correlation method\cite{a43,a44,a45}, big data mining, multiple gradual regression, and parallel multiple gradual regression. These methods have provided valuable insights into player performance by analyzing historical data and game statistics\cite{a46,a47,a48}.

With the advent of machine learning and deep learning techniques, researchers have increasingly turned to neural network models for more sophisticated analysis of tennis games\cite{a49}. These modern approaches offer the potential for more nuanced and accurate predictions by leveraging complex patterns and relationships within the data. Neural network models\cite{a50,a51}, particularly those based on deep learning architectures, have shown promise in capturing the dynamic and multi-faceted nature of tennis matches.

However, existing methods often focus on individual player performance without adequately considering the interaction between players from both sides. This limitation can lead to discrepancies in the evaluation results, as the performance of one player is inherently influenced by the actions and strategies of their opponent\cite{a52,a53,a54}. To address this gap, recent studies have introduced multi-level fuzzy comprehensive evaluation models to systematically assess in-game player performance. These models aim to capture the complex interplay between players and provide a more holistic view of game dynamics.

In addition to traditional statistical and machine learning methods, the concept of momentum has been introduced to quantify the winning trend and performance dynamics across a game. By selecting specific data indicators such as player streak, continuous player score, and score difference, researchers have been able to effectively capture these dynamics. This approach allows for a more nuanced understanding of how momentum shifts can influence the outcome of a match.

Recent advancements in generalized regression neural networks (GRNN) have also been applied to the prediction of tennis match outcomes\cite{a55,a56}. GRNN, as a parallel computing model, offers strong advantages in approximation ability, classification ability, and learning speed. However, the optimal spread value, which directly affects the prediction effect of the GRNN network, is typically determined using trial algorithms, which can be computationally complex and inefficient. To address this, our study introduces cross-validation to optimize the spread value of the GRNN, thereby improving prediction accuracy and reducing mean squared error (MSE).

In summary, our work builds on the foundation laid by traditional statistical methods, machine learning, and deep learning techniques, while introducing innovative approaches to better capture the complex dynamics of tennis matches. By incorporating multi-level fuzzy comprehensive evaluation models, momentum analysis, and optimized GRNN, our study advances the theoretical framework and offers practical tools for analysts and coaches in strategic game planning.

\section{Method}
\label{sec:Method}
In this section, we detail the methodological framework of our study, which is designed to enhance the predictive accuracy of tennis match outcomes and analyze player momentum.   We begin with the establishment of a Fuzzy Analytic Hierarchy Process (FAHP) model to evaluate player performance indicators, followed by the application of Principal Component Analysis (PCA) for data dimensionality reduction.   Subsequently, we construct a CV-GRNN model to predict match outcomes based on the reduced set of indicators.   We then perform a correlation verification to assess the relationship between the identified momentum indicators and match outcomes.   Finally, we refine our CV-GRNN model by incorporating additional statistically significant indicators, leading to improved predictive performance.

\begin{figure}[H]
  \centering
  \includegraphics[width=0.80\textwidth]{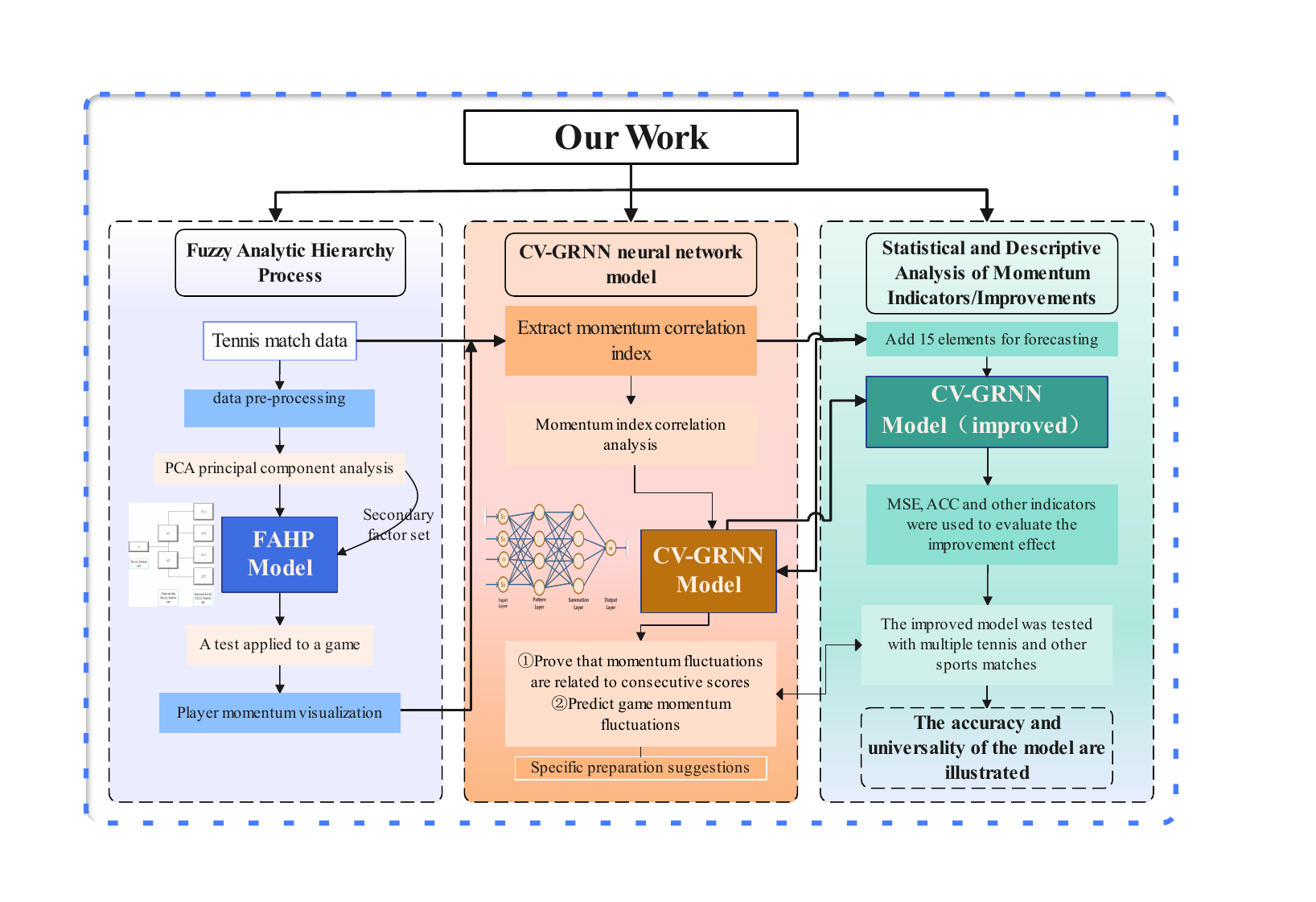} 
  \caption{The Overall Flow Chart of Techniques and Methods in this Paper}
  \label{flow}
\end{figure}

This section introduces technological approaches involved in this paper, while Figure \ref{flow} shows the overall process.

\subsection{The Establishment of The FAHP Model and Solution}

\subsubsection{Evaluation Index Establishment}

To describe player performance both scientifically and reasonably, this paper uses the sequence number of tennis matches as the classification standard. It also considers the time period completed by each sequence number as the division for carrying out statistical analysis of players in each match. For the first time, the following indicators have been selected:
\begin{itemize}[itemsep=0em, parsep=0em, topsep=0em, partopsep=0em]
    
    \item Number of wins $x_1$: The number of wins is one of the most basic indicators.

    \item Average winning time $x_2$: Winning time can reflect a player's staying power and endurance in the game. It is expressed as:
    \begin{equation}
        x_2=\frac{1}{n}\sum_{i=1}^{i=n}x_{2,i},
    \end{equation}
    This formula calculates the mean winning time over $n$ matches, providing insight into the player's endurance.
    \item Winning duration stability $x_3$: Winning duration stability can reveal the consistency of a player's performance across different matches.The expression for this is:
    \begin{equation}
        x_3=\frac{1}{n}\sum_{i=1}^{i=n}(x_{2,i}-x_{2,i-1}).
    \end{equation}
    This formula measures the variance in winning times, indicating how consistently a player performs.
    \item 
    Average score $x_4$ and total score $x_5$: These two indicators reflect the player's scoring ability and aggression. The expression is:
    \begin{equation}
        x_4=\frac{1}{m}\sum_{i=1}^{i=m}x_{4,i}\quad x_5=\sum_{i=1}^{i=m}x_{4,i},
    \end{equation}
    Here, $x_4$ represents the average score per match, while $x_5$ is the total score accumulated over $m$ matches.
    \item High scoring rate $x_6$: High scoring rate reflects players' performance in key moments.  In this article, $m'$ represents the number of times the score is 40 or higher. The calculation is as follows:
    \begin{equation}
        x_6=\frac{m'}{m} \times  100\%.
    \end{equation}
    This formula calculates the percentage of high-scoring instances, indicating clutch performance.
    \item Let $x_{7i}$ represent the number of points earned in the first $i$match, and $x_{all,i}$ represent the total number of points earned in the first $i$ match. The expression is as follows:
    \begin{equation}
        x_7=\frac{1}{m}\sum_{i=1}^{i=m}\frac{x_{7,i}}{x_{all,i}},
        \qquad x_8=\frac{1}{m}\sum_{i=1}^{i=m}(x_{7,i}-x_7)^2.
    \end{equation}
    These formulas measure the average points earned and the consistency of point scoring across matches.
     \item Serve score $x_9$ and second serve score $x_{10}$: These two metrics reflect a player's serve ability. They are calculated as the sum of the points earned from first and second serves, respectively.
     
     \item  First serve score rate $x_{11}$ and second serve score rate $x_{12}$: Two indicators further measure the effectiveness of a player's serve.  The formula is provided below:
     \begin{equation}
         x_{11}=\frac{x_{9}}{x_9+x_{10}}\quad  \qquad  x_{12}=\frac{x_{10}}{x_9+x_{10}}.
     \end{equation}
     These ratios indicate the success rates of first and second serves.
     \item ACE number $x_{13}$: Reflects the serve power and skill level of the player. This article directly uses the given data values.
     \item Average win rate $x_{14}$: Reflects the player's win rate performance in different matches. A higher average win rate indicates that the player has a strong match state and competitive level. The formula is:
     \begin{equation}
         x_{14}=\frac{1}{m}\sum_{i=1}^{i=m}x_{14,i}\times  100\%.
     \end{equation}
     This formula calculates the average win rate as a percentage.
     \item Hit a non-trigger ball rate $x_{15}$: Reflects a player's explosive power and potential in the game. The formula is:\begin{equation}
         x_{15}=\frac{1}{m}\sum_{i=1}^{i=m}x_{15,i}  \times  100\%.
     \end{equation}
     This rate measures the frequency of hitting non-trigger balls, indicating aggressive play.
     \item  Miss two serves and lose points $x_{16}$and make unforced errors $x_{17}$: Reflect the player's errors during the match. The expressions are as follows:
     \begin{equation}
         x_{16}=\frac{1}{m}\sum_{i=1}^{i=m}x_{16,i}  \times  100\% \qquad \quad x_{17}=\frac{1}{m}\sum_{i=1}^{i=m}x_{17,i}  \times  100\%.
     \end{equation}
     These formulas measure the frequency of serve errors and unforced errors.
     \item Net success rate $x_{18}$and net win rate $x_{19}$: Reflect the player's ability and effectiveness in taking the initiative to go to the net during the match. The expressions are as follows:
     \begin{equation}
         x_{18}=\frac{1}{m}\sum_{i=1}^{i=m}x_{18,i}  \times  100\% \qquad \quad x_{19}=\frac{1}{m}\sum_{i=1}^{i=m}x_{19,i}  \times  100\%.
     \end{equation}
     These rates measure the success and win rates when approaching the net.
     \item Missed chances to win an opponent's serve $x_{20}$: Reflects the number of times a player misses a chance to win an opponent's serve during a match. The formula is:
          \begin{equation}
         x_{20}=\frac{1}{m}\sum_{i=1}^{i=m}x_{20,i}  \times  100\%.
     \end{equation}
   This rate measures the frequency of missed opportunities to break the opponent's serve.
     \item    Average run distance $x_{21}$and run distance stability $x_{22}$: Reflect a player's physical fitness and running status during the game. The expressions are as follows:
     \begin{equation}
         x_{21}=\frac{1}{m}\sum_{i=1}^{i=m}x_{21,i}, \qquad x_{22}=\frac{1}{m}\sum_{i=1}^{i=m}(x_{22}-x_{22,i})^2.
     \end{equation}
     These metrics measure the average running distance and the stability of running performance.
\end{itemize}

To ensure that the indicators selected in this paper contribute significantly to the competition, the PCA (Principal Component Analysis) dimensionality reduction method  \cite{ref1} is used to effectively capture key information in the data. The dimensions of 21 indicators are simplified into 10 principal components.

\textbf{The Establishment  of Evaluation Model}

Given the complexity of the problem with many factors and fuzzy overall evaluation criteria, the fuzzy comprehensive evaluation method is chosen to establish the evaluation model \cite{ref2}. Due to the numerous indicators, the indicators are stratified, and a two-level fuzzy comprehensive evaluation system is employed. The secondary factor set includes the eleven indicators selected in the first step, and a primary factor set is established to classify these secondary factors.
\begin{enumerate}[itemsep=0em, parsep=0em, topsep=0em, partopsep=0em]
    \item First, set the first first-level factor $A_1$to investigate the athlete's physical fitness, we consider players' average running distance $x_{21}$ and players' running explosion $x_{22}$, which correspond to the secondary factor set $A_1^1,A_1^2$in the second-level factor set $A_1$of the first-level factor set.

Namely: $A_1\textbf{(physical fitness)}=\{A_1^1,A_1^2\}$.Where$A_1^1,A_1^2$ are extremely significant indicators.
    \item Next, the second level factor $A_2$ can be set to examine the players' serving scoring ability, this includes players' scoring on the first serve $x_9$, scoring on the second serve $x_{10}$, the first serve score rate $x_{11}$, and the second serve score rate $x_{12}$, corresponding level factors set $A_2^2$ concentration of secondary factors $A_2^1,A_2^2,A_2^3,A_2^4$.
    
    The $A_2\textbf{(serving proficiency)}=\{  A_2^1,A_2^2,A_2^3,A_2^4 \}$, where $A_2^1,A_2^2,A_2^3,A_2^4$ are very large.
    \item    Then, the third first-level factor $A_3$ is set to examine the winning strength of the players, this includes the number of wins $x_1$ and the average winning time $x_2$, corresponding to the secondary factor set $A_3^1,A_3^2$ in the first-level factor set $A_3$.
    The $A_3\textbf{(winning capability)}=A_3^1,A_3^2$, where $A_3^1$ for extremely large index, $A_3^2$ for very small target.
    \item
    Finally, a fourth first-level factor $A_4$ is set to examine the comprehensive scoring ability of the players, this includes the total score $x_5$, the average score $x_4$, and the average proportion of players' scores $x_7$, which correspond to the secondary factors $A_4^1,A_4^2, and A_4^3$ in the first-level factor set $A_4$.
    The $A_4\textbf{(overall score)}=\{ A_4^1,A_4^2,A_4^3  \}$, where $A_4^1,A_4^2,A_4^3$ are significant indicators.
\end{enumerate}

\begin{figure}[H]
  \centering
  \includegraphics[width=0.75\textwidth]{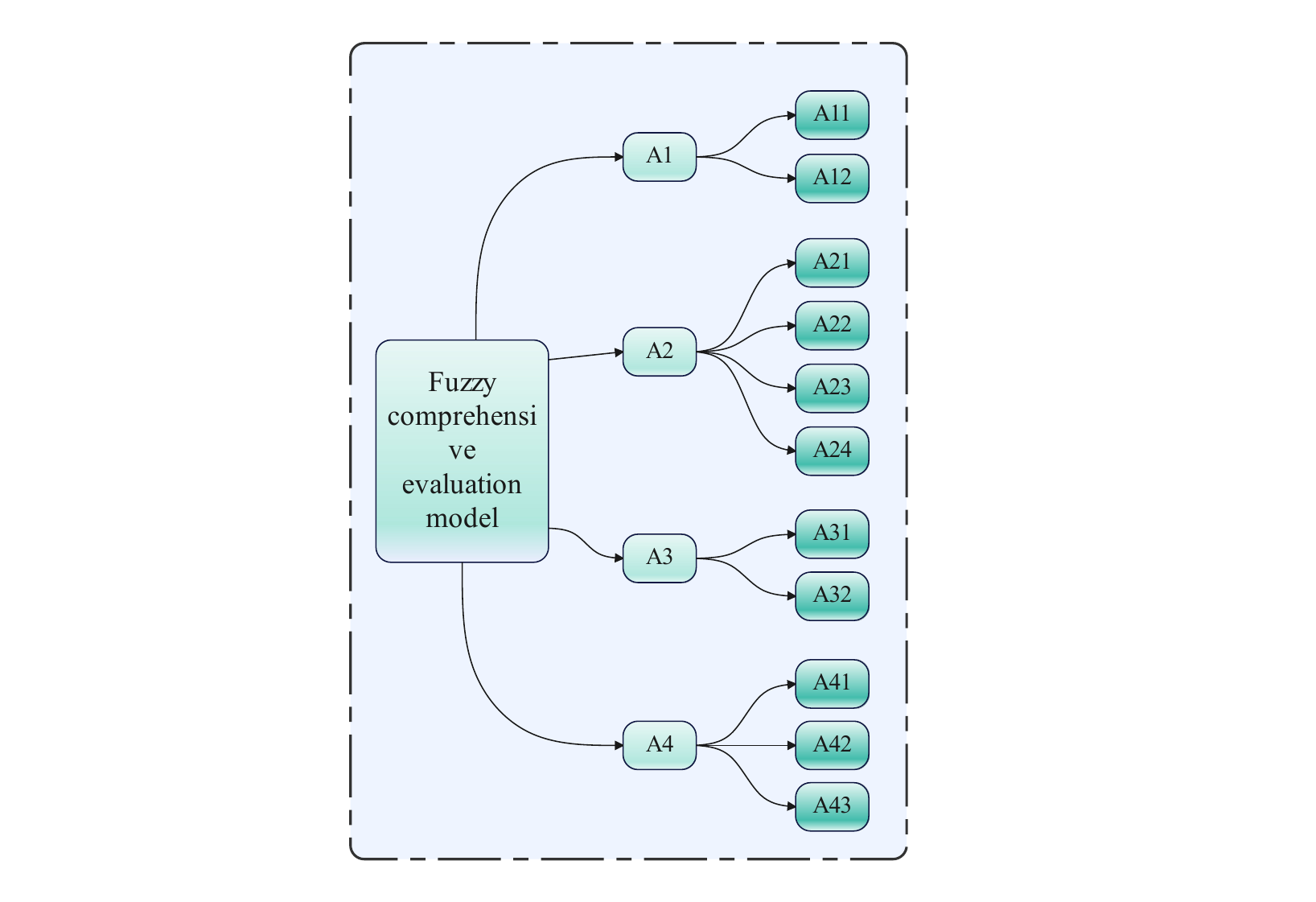} 
  \caption{Schematic Diagram of Second-Level Fuzzy Comprehensive Evaluation System Model}
  \label{MCM20P3}
\end{figure}
The fundamental framework of the evaluation model has been established, as shown in Figure \ref{MCM20P3}.



\textbf{Positive and normalized:} 
In the evaluation model, it is crucial to standardize the data to ensure that each indicator contributes proportionally to the final evaluation.   For indicators that are inherently small but significantly influential, such as the average winning time ($x_2$), a positive transformation is applied to ensure they align with other indicators in a positive framework.

For the very small indicator $A_3^2$, it is forward processed, in addition, other indicators can also be normalized. Among them, the forward formula is as follows:
\begin{equation}
    x_p=\frac{x_{max}-x}{x_{max}-x_{min}},
\end{equation}
\textbf{Set comments:}
To quantify the momentum of a player, a set of evaluation grades ($V$) is defined, capturing the spectrum of performance from very weak to very strong.
\begin{equation}
    V=\{ Very weak,Weak,Weaker,Moderate,Stronger, Strong,Very strong\}
    \label{very12}
\end{equation}

Equation (\ref{very12}) categorizes the qualitative assessment of player momentum into quantifiable levels, allowing for a more nuanced analysis.

The momentum score is calculated by assigning different weights to each level of the evaluation set ($V$), reflecting the significance of each level in the overall performance assessment..Combined with relevant data, the formula for calculating momentum score is set as:
\begin{equation}
Score\,\,=\,\,10V\left( 1 \right) +30 V\left( 2 \right) +40 V\left( 3 \right) +60 V\left( 4 \right) +70 V\left( 5 \right) +80V\left( 6 \right) +100V\left( 7 \right) 
\label{4070v}
\end{equation}
Equation (\ref{4070v}) translates the qualitative momentum categories into a quantitative score, facilitating a more objective comparison of player performance.

With reference to relevant literature, the centralization weight of primary factor is set, and the entropy weight method is adopted to give weight to the secondary factor set  \cite{ref3}.

First, the probability in the relative entropy calculation is computed by taking the proportion of the \(i\)th sample of the \(j\)th index as the probability in the calculation of relative entropy. A probability matrix \(P\) is established, where the calculation formula for each element \(p_{ij}\) in \(P\) is as follows:

\begin{equation}
p_{ij}=\frac{z_{ij}}{\sum_{i=1}^n{z_{ij}}}
\end{equation}

In the above formula, the sum of the probabilities corresponding to each indicator is 1.

Next, the information entropy $e_{j}$of each index is calculated and the information utility value $d_{j}$is further calculated as follows:
\begin{equation}
e_j=-\frac{1}{\ln\mathrm{n}}\sum_{\mathrm{i}=1}^{\mathrm{n}}{\mathrm{p}_{\mathrm{ij}}\ln\mathrm{(p}_{\mathrm{ij}})\mathrm{(j}=1,2)} ,\quad 
d_j=1-e_j
\end{equation}
Finally, the information utility value is normalized to obtain the entropy weight of the index, which is then taken as the weight of the index.

\begin{small}
 \begin{equation}
W_{2}^{j}=d{{_j}\Bigg/{\sum_{j=1}^2{d_j}}}\mathrm{}
\label{w2j}
\end{equation}
\end{small}

Equation (\ref{w2j}) normalizes the information utility value to obtain the entropy weight of each index, ensuring a balanced contribution to the overall evaluation.

The weight sets $K_2^1,K_2^2,K_2^3,K_2^4$corresponding to the second-order factor set are obtained respectively.

\textbf{Define membership functions:} 
For comment set V $\{Weak,Weaker,Moderate,Stronger, Strong\}$for middle-type comments, and \ {Very Weak \}, \ {Very Strong \}for extreme-type comments, partial comments are classified as slightly small or slightly large. Accordingly, the assignment method is used to determine the membership function for each index corresponding to the review set.

\textbf{Calculate the judging vector:}
Based on the determined membership function, the evaluation vector $R_i=A(u_i)$ for each index is calculated for$\quad i=1,2,3,4,5,6,7$.

\textbf{First-level membership set:}
For each child separately level fuzzy comprehensive evaluation factor set, namely:
\textbf{Second-level membership set:}
$B=A*R.$

\subsubsection{The Evaluation Model Sloving:}
A match between players Carlos Alcaraz and Nicolas Jarry with match-ID '2023-wimbledon-1301' was used to validate our evaluation model. During this process, we substituted the performance status data of the two players at different times for calculation. A model can be built that describes the momentum of the players in the match.

\textbf{Step 1:}
We extracted the relevant data of the match and calculated the index data for 11 secondary factor sets related to the players. The formula has been shown above, so we will not repeat it here.

\textbf{Step 2:}
The data are processed in a positive and standardized way.

\textbf{Step 3:}
In the construction of the fuzzy comprehensive evaluation model, it is essential to assign appropriate weights to different factors to reflect their relative importance in the overall assessment.  Based on a thorough review of relevant literature and previous studies, we have determined the centralized weights for the first-level factors as follows:

\begin{equation}
    A=[0.15,0.25,0.35,0.25]
    \label{015A}
\end{equation}
Equation (\ref{015A}) represents the centralized weights assigned to the first-level factors after a comprehensive review of existing literature. These weights are crucial in capturing the relative significance of each factor in influencing the player's performance.

The choice of weights is based on the analysis of previous studies that have identified the impact of different factors on player performance. The weights are designed to ensure that no single factor dominates the evaluation, thereby maintaining a balanced and comprehensive assessment.

\textbf{Step 4:} To effectively convert raw competition data into a meaningful evaluation of athletes' performance, it is necessary to define a membership function.  This function maps the quantitative data into qualitative categories, allowing for a more nuanced analysis of the athletes' performance dynamics.  The membership function is designed as follows:
\begin{small}
\begin{equation}
\label{eq:membership-function}
\left\{
\begin{aligned}
R_1 & = (0 \leqslant U < 0.05) + \left(\frac{0.065 - U}{0.015}\right) \cdot (U \geqslant 0.05 \land U < 0.065); \\
R_2 & = \left(\frac{U - 0.06}{0.1}\right) \cdot (0.06 \leqslant U < 0.16) + (0.16 \leqslant U < 0.3) + \left(\frac{0.35 - U}{0.05}\right) \cdot (0.3 \leqslant U < 0.35); \\
R_3 & = \left(\frac{U - 0.25}{0.05}\right) \cdot (0.25 \leqslant U < 0.3) + (0.3 \leqslant U < 0.35) + \left(\frac{0.4 - U}{0.05}\right) \cdot (0.35 \leqslant U < 0.4); \\
R_4 & = \left(\frac{U - 0.25}{0.15}\right) \cdot (0.25 \leqslant U < 0.4) + (0.4 \leqslant U < 0.6) + \left(\frac{0.75 - U}{0.15}\right) \cdot (0.6 \leqslant U < 0.75); \\
R_5 & = \left(\frac{0.7 - U}{0.1}\right) \cdot (0.6 \leqslant U < 0.7) + (0.55 \leqslant U < 0.6) + \left(\frac{U - 0.5}{0.1}\right) \cdot (0.5 \leqslant U < 0.52); \\
R_6 & = \left(\frac{0.9 - U}{0.06}\right) \cdot (0.84 \leqslant U < 0.9) + (0.7 \leqslant U < 0.84) + \left(\frac{U - 0.65}{0.05}\right) \cdot (0.65 \leqslant U < 0.7); \\
R_7 & = (0.8 \leqslant U < 1) + \left(\frac{U - 0.75}{0.05}\right) \cdot (0.75 \leqslant U < 0.8); \\
\end{aligned}
\right.
\end{equation}
\end{small}

Equation (\ref{eq:membership-function}) defines the membership functions that map the quantitative performance data into qualitative categories, allowing for a detailed analysis of the athletes' performance dynamics.

To provide a visual representation of the membership function, a graph is plotted as shown in Figure \ref{member}.  This visual aid helps in understanding the relationship between the quantitative data and the qualitative categories defined by the membership function.

\begin{figure}[H]
  \centering
  \includegraphics[width=0.65\textwidth]{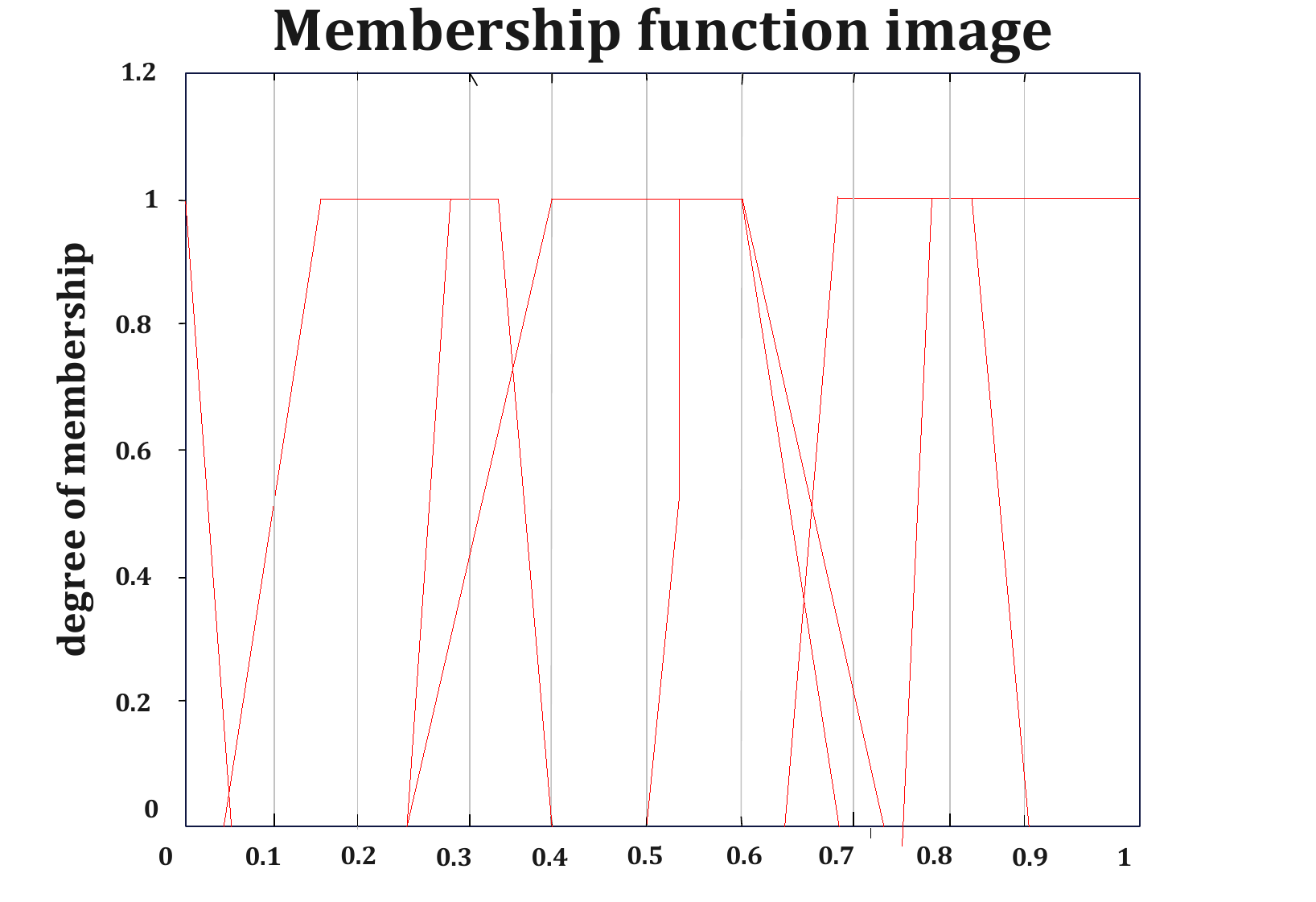} 
  \caption{The Membership Function Graph Plotted According to Membership Function}
  \label{member}
\end{figure}

Figure \ref{member} provides a graphical representation of the membership function, illustrating how the quantitative performance data is translated into qualitative categories.  This visualization is crucial for understanding the dynamics of athletes' performance and their competitive state during the match.

\textbf{Step 5:}
The evaluation vector corresponding to the index is calculated, and the first level fuzzy comprehensive evaluation is carried out for each sub-factor set.

\textbf{Step 6:}
The first and second level membership sets are calculated successively.

\textbf{Step 7:}
The corresponding membership degree of the weight set is calculated according to the second-level membership degree set, and the evaluation images of the momentum of different players at different moments are drawn, so as to visually describe the change and fluctuation of the momentum of players in the competition process (see Figure \ref{fig:momentum}).


\begin{figure}[H]
  \centering
  \includegraphics[width=0.65\textwidth]{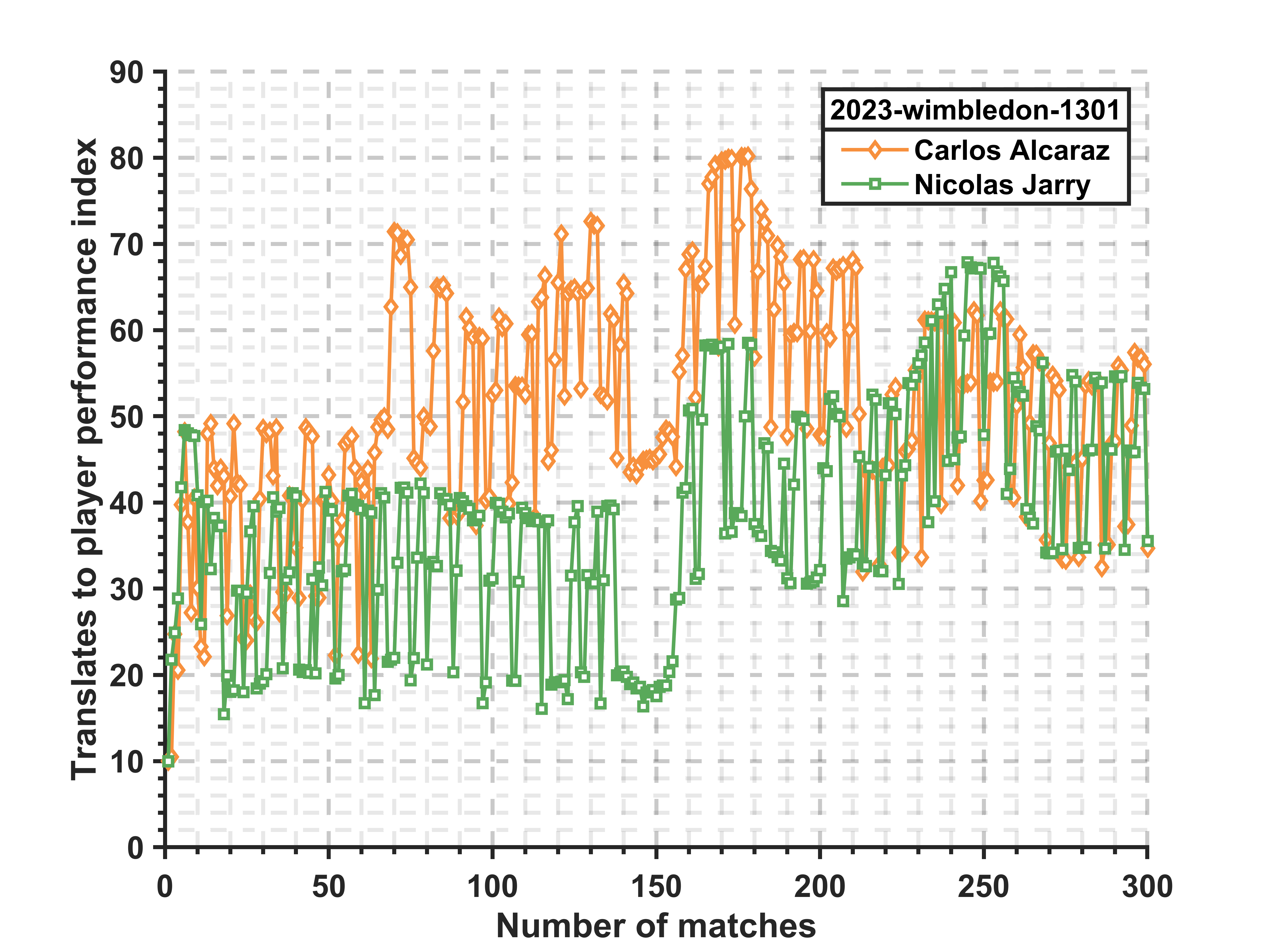} 
  \caption{The Momentum of The Two Players Varies at Different Times}
  \label{fig:momentum}
\end{figure}

\textbf{Step 8:}
After solving the model, it is evident from the images that, during the specified race process, Player 1's momentum is dominant in the early and middle stages, while Player 2's momentum equals that in the late stage, aligning with the actual development of the race. At this point, we have constructed a comprehensive evaluation system and have applied it to a specific match to test and determine which player is performing better at any given moment. Additionally, we have plotted the momentum scores of the players at different moments to provide a visualization based on the evaluation model we have constructed. This clearly describes the development of player momentum during the game.

\subsection{Correlation Verification and CV-GRNN }
After initial data preprocessing, momentum indicators tied to match swings and players' consistent victories are distilled.  These indicators are then quantified and processed to determine players' likelihood of success in future games.  Correlation analysis is employed to assess the relationship between these indicators and player outcomes.  Utilizing the CV-GRNN neural network model, match fluctuations are anticipated, and model accuracy is appraised through error analysis.  Finally, statistical examination of momentum indices during successful player transitions guides tailored recommendations for players.

\subsection{Correlation Verification}

During the game, the number of consecutive wins, the score difference between players and opponents, the number of consecutive scores, and the point difference between players and opponents can be used as continuous indicators to affect the players' game results, ultimately defining the "next win" status indicator.

\begin{itemize}[itemsep=0em, parsep=0em, topsep=0em, partopsep=0em]
\item \textbf{Player streak $S_1$:} Indicates the number of consecutive wins a player has in a match. This can be directly calculated by counting 'p-sets'.
\item \textbf{Player-opponent score difference $s_2$:} Represents the difference between a player's score and the opponent's score.   The calculation formula is:
$ S_2=p_{1-score}-p_{2-score}$.

\item \textbf{Number of consecutive points scored $S_3$:} Indicates the number of consecutive points scored by a player in a match. The statistical method is as follows: Each time a player wins a streak, the value is increased by one. If the streak is interrupted, the value resets to 0.

\item \textbf{Player and opponent point difference $S_4$:} Indicates the difference between the points won by a player and those won by the opponent.   The formula is: $S_4=p_{1,points-won}-p_{2,points-won}$.

\item \textbf{Next win status indicator $\omega$:}
Uses the 'point-victor' data to make a judgment. If the next score is a study object, it is recorded as 1; If not, it is recorded as 0.
\end{itemize}

To enhance the predictive accuracy of our model, it is essential to understand how each momentum indicator influences the outcome of subsequent matches.  For this purpose, we introduce the Pearson correlation coefficient ($R$), a statistical measure that quantifies the linear relationship between two variables, expressing both the strength and direction of the association. It provides a standardized approach to evaluate the relationship between each momentum indicator and the next match outcome, allowing for a quantitative assessment of their interdependence.

The Pearson correlation coefficient is calculated using the following formula, which considers the deviations of each data point from the mean and emphasizes the co-variability of the indicators:

\begin{equation}
    R = \begin{cases} 
    \frac{\sum{(S_i - \bar{S_i})(S_j - \bar{S_j})}}{\sqrt{\sum{(S_i - \bar{S})^2} \cdot \sum{(S_j - \bar{S_j})^2}}} & \text{\small When all indicators are momentum-based} \\  
    \frac{\sum{(S_i - \bar{S_i})(\omega - \bar{\omega})}}{\sqrt{\sum{(S_i - \bar{S})^2} \cdot \sum{(\omega - \bar{\omega})^2}}} & \qquad \qquad \qquad  \qquad \text{\small Else.}
    \label{MATCH}
\end{cases}
\end{equation}

Equation (\ref{MATCH}) presents the conditional formula for calculating the Pearson correlation coefficient.  When both variables are momentum-based indicators, the first formula applies, focusing on the relationship between pairs of indicators.  If, however, the comparison involves a momentum indicator and the next match outcome ($\omega$), the second formula is used to capture their direct correlation.

\subsection{Establishment and Solution of CV-GRNN Model}

The Generalized Regression Neural Network model (GRNN) belongs to the radial basis neural network models \cite{ref4}, which have strong nonlinear mapping capabilities and robustness. 

However, the traditional GRNN model has the disadvantages of slow convergence and susceptibility to producing locally optimal solutions. In this paper, the smoothing factor and radial basis expansion rate of the GRNN model are studied based on the principle of cross-validation to optimize and improve, finally building the CV-GRNN model \cite{ref5}.

\subsubsection{Generalized Regression Neural Network GRNN}

The Gaussian Radial Basis Function Neural Network (GRNN) is recognized for its robust regression capabilities and is ideally suited for predictive analytics in sports performance modeling.  In this paper, we employ a GRNN with a specific focus on tennis match outcomes.  The network architecture comprises four primary layers: the input layer, the radial basis function layer, the summation layer, and the output layer.  Each of these layers plays a crucial role in processing the input data and generating accurate predictions for tennis match outcomes.

\begin{enumerate}
    \item Input Layer: Incorporates key performance indicators that significantly influence the match dynamics. These include 'Player Win Streak ($S_1$)', 'Score Difference ($S_2$)', 'Consecutive Scores ($S_3$)', and 'Point Difference ($S_4$)'.
    \item Pattern Layer: Processes the input data, assigning each input to a radial basis function.
    \item Summation Layer: Aggregates the outputs from the pattern layer to prepare for the final output stage.
    \item Output Layer: Provides the predicted outcome, specifically the 'Next Win Indicator ($\omega$)', which predicts the likelihood of a player's victory in subsequent matches.
\end{enumerate}

Figure  \ref{fig:example}  illustrates the model architecture, providing a visual representation of the data flow and transformation within the GRNN.

\begin{figure}[H]
  \centering
  \includegraphics[width=0.55\textwidth]{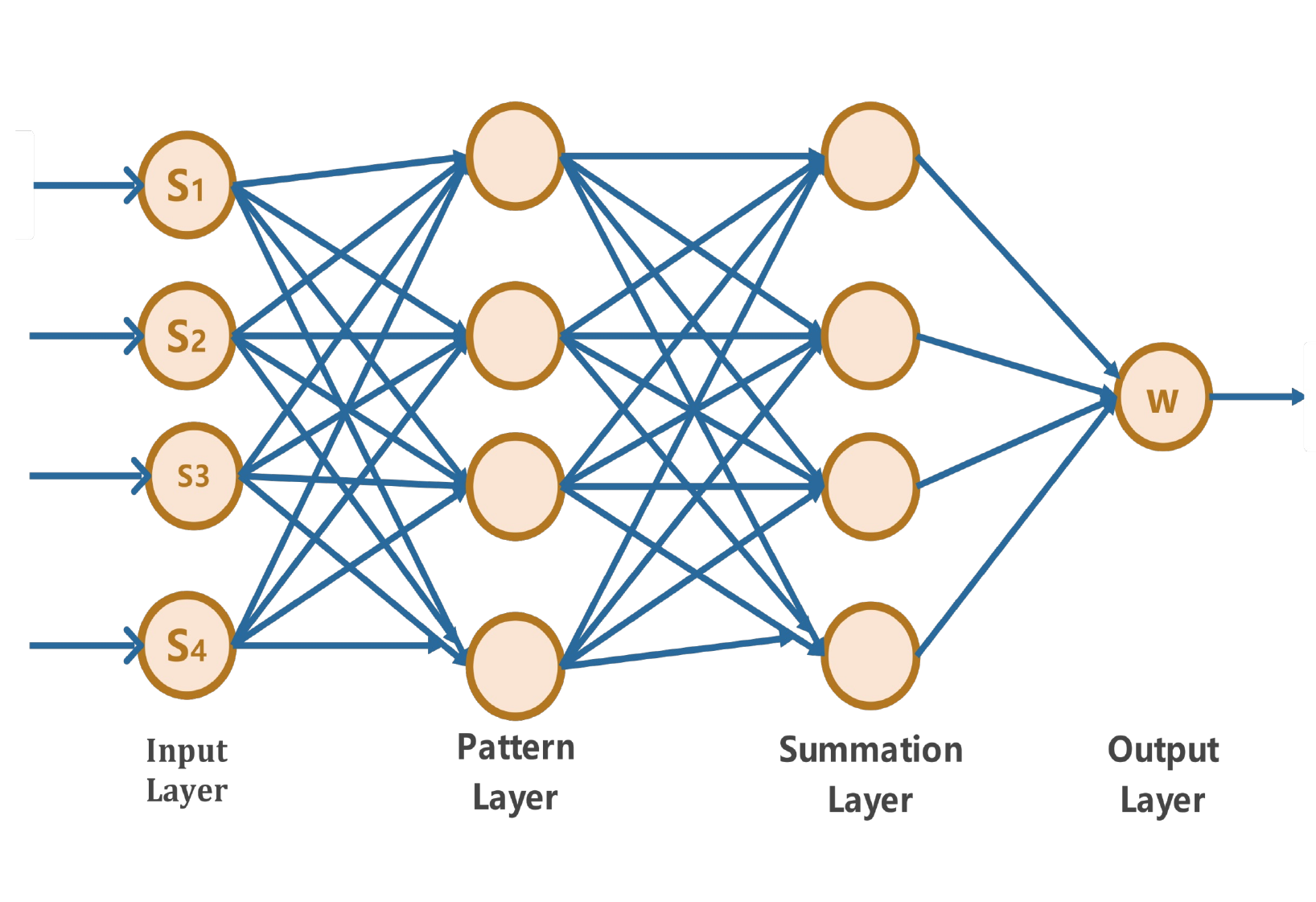} 
  \caption{Generalized Regression Neural Network Model Structure}
  \label{fig:example}
\end{figure}

For the GRNN, only one smoothing factor,$\sigma$, needs adjustment. To overcome the defects of the traditional $\sigma$ optimization method and improve prediction accuracy, a cross-validation (CV) algorithm is introduced for optimization.  Thus solving  \cite{ref6}.

Next, in order to better evaluate the advantages and disadvantages of the traditional model and the CV-GRNN model used in this paper, the mean square error (MSE) and accuracy rate (ACC) are adopted as the performance evaluation indexes of the prediction model. The formula is as follows:
\begin{equation}
    MSE= \frac{1}{n} \sum_{i=1}^{n} (\omega_i - \hat{\omega_i})^2    \
    \label{MSEACC}
\end{equation}
\begin{equation}
    ACC = \frac{\text{Correct Predictions}}{\text{Total Samples}} \times 100\%
    \label{MSEACC2}
\end{equation}
where $n$ is the number of samples, $\omega_i$is the actual value, and $\hat{\omega_i}$is the predicted value.

After organizing the established process, the model prediction process of CV-GRNN is obtained as Algorithm \ref{alg:cv-grnn-prediction} describes.The CV-GRNN prediction process is outlined in Algorithm 1. This algorithm takes into account the influencing factors ($S_1, S_2, S_3, S_4$) and the output factor ($\omega$) to predict the outcome of a tennis match.

\begin{algorithm}[H]
\small
\caption{CV-GRNN Prediction}
\label{alg:cv-grnn-prediction}

\textbf{Input:} $S_1, S_2, S_3, S_4$ (influencing factors), $\omega$ (output factor)

\textbf{Output:} Predicted outcome $\hat{\omega}$

\begin{algorithmic}[1]

\State Establish a database with $S_1, S_2, S_3, S_4$ and $\omega$.

\State Normalize sample data to prevent GRNN convergence issues.

\State Split data into training and prediction sets.

\State Optimize smoothing factor $\sigma$ using cross-validation.

\State Build a GRNN network with optimal $\sigma$.

\State Predict player wins in the next match using $\omega$.

\State Iterate until convergence or target accuracy is achieved, and denormalize prediction results.

\State Output predicted outcome $\hat{\omega}$ and evaluate model performance.
\end{algorithmic}
\end{algorithm}

\section{EXPERIMENTS AND ANALYSES}
\label{EXPERIMENTS AND ANALYSES}

In this section, we present the experimental design and analytical processes employed to validate the proposed model.    Our primary objective was to assess the effectiveness and accuracy of integrating a fuzzy comprehensive evaluation model with a Cascading Validation-General Regression Neural Network (CV-GRNN) in predicting tennis match outcomes.    The following steps were undertaken:

1.    Data Collection: Historical data from various international tennis tournaments, including Wimbledon, were gathered.    This dataset included critical match statistics such as player win streaks, score differences, consecutive scores, and point differences.

2.    Data Preprocessing: The collected data underwent cleaning and normalization to ensure high quality and to eliminate noise that could affect model predictions.

3.    Feature Selection: Principal Component Analysis (PCA) was utilized to reduce the dimensionality of the data and identify key statistical indicators that significantly impact match outcomes.

4.    Model Construction: Based on the preprocessed data and selected features, we constructed the fuzzy comprehensive evaluation model and the CV-GRNN model.    This involved parameter determination, training, and validation phases.

5.    Performance Evaluation: The model's predictive performance was evaluated using metrics such as Mean Squared Error (MSE) and accuracy (ACC).

6.    Results Analysis: A detailed analysis of the model's predictions was conducted to determine its stability and generalizability across different match conditions.

Through this comprehensive process, we aim to demonstrate the superiority of our proposed method in predicting tennis match outcomes and explore its potential applications in sports analytics.

\subsection{Data Sources}

The data employed in this study was exclusively extracted from the "Wimbledon 2023 Gentlemen's singles matches after the second round" dataset.  This dataset provides a granular level of detail, capturing every scoring point throughout the tennis matches.  The data is predominantly structured, encompassing a range of information such as player scores, faults, and match durations.Data characteristics and types are as follows:

\begin{itemize}
    \item \textbf{Match ID:}A unique identifier for each match, formatted as "2023-wimbledon-1701," which denotes the first match in the seventh round of the 2023 Wimbledon Championship.
    \item \textbf{Player 1 \& Player 2: }The names of the competing players, e.g., "Carlos Alcaraz" and "Novak Djokovic."
    \item \textbf{Elapsed Time:}The time elapsed since the match's commencement, recorded in minutes and seconds (e.g., "0:01:31" indicates one minute and 31 seconds into the game).
    \item \textbf{Set No: }Indicates the current set number, with "3" signifying that three sets have been won out of a best-of-five sets format.
    \item \textbf{Game No:}Represents the current game number within the set, with "1" denoting the first game.
    \item \textbf{Point No: }The sequence of points within a game, with "12" marking the twelfth point.
    \item \textbf{P1 Sets \& P2 Sets:} The number of sets won by Player 1 and Player 2, respectively, with "2" indicating two sets won.
    \item \textbf{P1 Games \& P2 Games:} The number of games won by Player 1 and Player 2, respectively, with "6" indicating six games won.
\end{itemize}

\subsection{Data Preprocessing}

\subsubsection{Testing and Handling Missing Values}

In the realm of sports analytics, particularly tennis match analysis, datasets can often present missing values due to various reasons such as data collection errors, incomplete records, or the inherent unpredictability of match conditions. The Wimbledon 2023 Gentlemen's singles matches after the second round dataset is not exempt from this challenge. Addressing missing values is crucial as they can introduce bias, reduce the sample size, and affect the reliability and validity of the analytical outcomes.

The initial step in our data preprocessing phase was to assess the missing rate of various statistical indicators. As demonstrated in Table \ref{missingrate}, different motion parameters such as Speed Mph, Serve Width, Serve Depth, and Return Depth have varying missing rates, with the most significant being Return Depth at 0.1797\%.

\definecolor{Snuff}{rgb}{1,1,1}
\begin{table}[H]
\centering
\caption{Missing Rate Table}
\begin{tblr}{
  cells = {c},	
  row{1} = {Snuff},
  hlines,
}
\textbf{Motion Parameters}  & \textbf{Speed Mph} & \textbf{Serve Width} & \textbf{Serve Depth} & \textbf{Return Depth} \\
\textbf{Missing Percentage} & 0.1032             & 0.0074               & 0.0074               & 0.1797                
\label{missingrate}
\end{tblr}
\end{table}

\subsubsection{Importance of Addressing Missing Values}

The accurate prediction of tennis match outcomes relies heavily on the integrity and completeness of the dataset. Missing values can lead to several issues:
\begin{itemize}
    \item Biased Analysis: The exclusion of data points can lead to a biased representation of the dataset, potentially skewing the analysis towards the available data.
    \item Reduced Sample Size: Missing values can reduce the sample size, thereby affecting the statistical power of the analysis and the generalizability of the results.
    \item Impact on Model Performance: In machine learning models, incomplete data can hinder the model's ability to learn patterns, thus affecting its predictive performance.
\end{itemize}

\subsubsection{Methodology for Handling Missing Values}

Given the importance of each player's data in drawing comprehensive conclusions, discarding entire rows of data was not a viable option. Instead, we treated each row of data as a vector. For a row $i$, the vector is represented as  \(A_i=[a_{i1},a_{i2},\cdots,a_{im}]\),where 
$m$ is the total number of columns in the dataset. This approach allowed us to systematically address missing values.

To measure the similarity between data vectors and effectively handle missing values, we introduced the Euclidean distance. This metric is pivotal in calculating the straight-line distance between two vectors in an 
n-dimensional space, providing a measure of similarity that is crucial for imputing missing values. The Euclidean distance $d$  between two points \(A_i\) and \(A_j\) is calculated as follows:

\begin{equation}
\label{eq:01}
    d(A_i, A_j) = \sqrt{\sum_{k=1}^{m} (a_{ik} - a_{jk})^2} 
\end{equation}

Assuming that \(A_i\) is a vector with missing value indices, this paper uses the corresponding indices in \(A_j\) of the minimum \(d(A_i,A_j)\) to replace the missing values.

This method aids in identifying the proximity between different data points, facilitating the process of imputing missing values based on similar, complete records. By leveraging the Euclidean distance, we ensure a robust and accurate approach to handling missing data, thereby bolstering the reliability of our analytical outcomes.

\subsubsection{Test of Outliers}

Next, this paper examines the data structure in depth and finds character information, "AD," in the 'p1-score' and 'p2-score' columns. After consulting relevant literature, "AD" is understood to refer to "advantage," a term used in tennis matches to indicate that a player has taken the lead when competing for match points. To simplify the model, "AD" is uniformly converted to 55 in this analysis.

To further enhance the robustness and accuracy of the model, a separate box method is employed to process continuous data. A box plot was used to visualize the results of the box divisions, as shown in Figure \ref{Boxplot}.

\begin{figure}[H]
  \centering
  \includegraphics[width=0.65\textwidth]{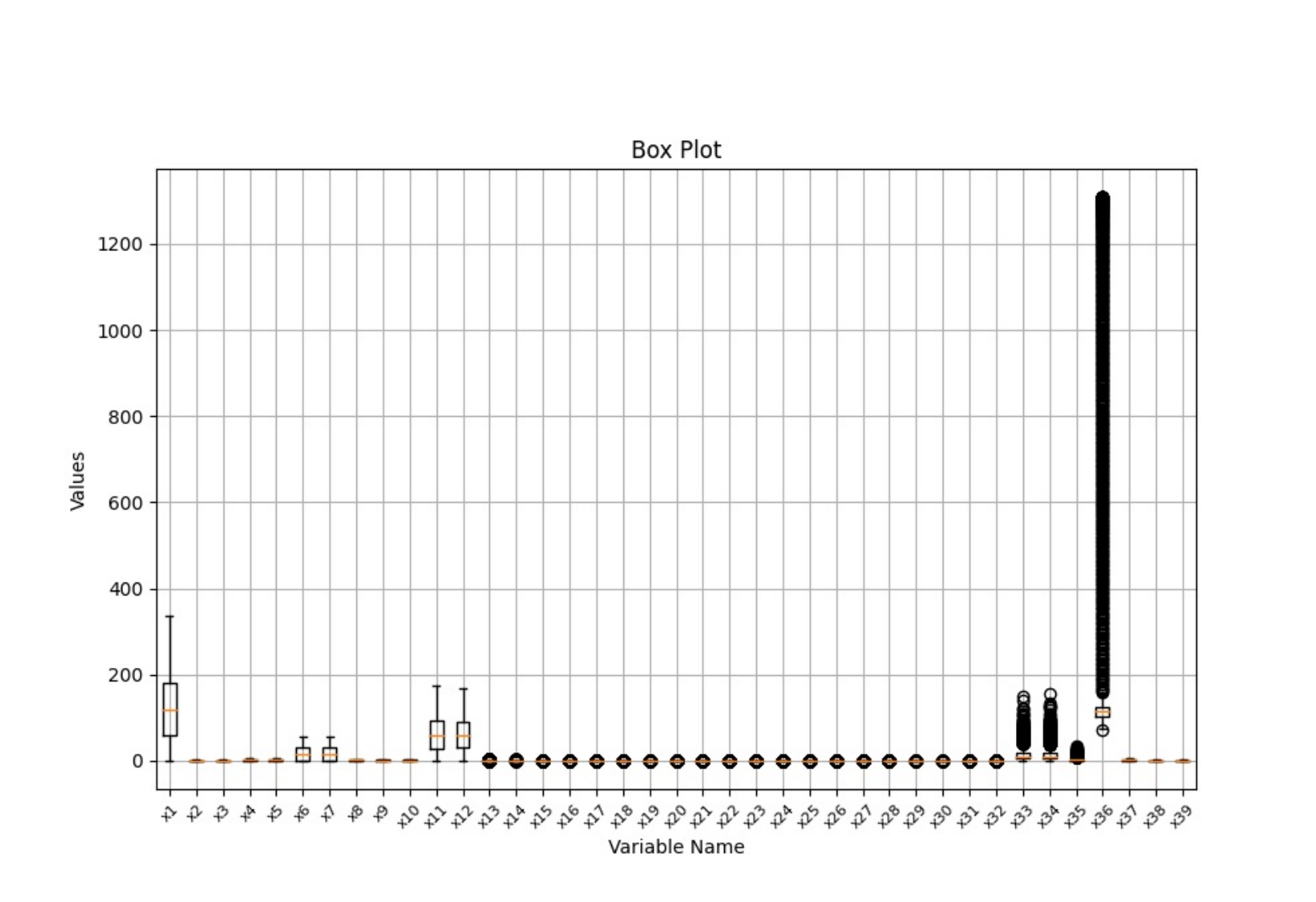} 
  \caption{The Box Plot Analysis}
  \label{Boxplot}
\end{figure}

Analyzing the boxplot, 'speed-mph' exhibits the most outliers, followed by 'distance-run'. The paper refers to data showing a maximum serving speed of 141 miles per hour. Running length fluctuation correlates with game intensity and other factors. Therefore, all data are considered within an acceptable range, eliminating the need for outlier removal.

The careful handling of missing values is essential for maintaining the robustness of our analytical model. By employing a strategy that considers the proximity of data points, we can effectively impute missing values and ensure that our model training and evaluation reflect a comprehensive understanding of the data. This approach stands as a testament to the meticulous data preprocessing required for high-fidelity sports analytics and further solidifies the foundation for our predictive model's accuracy and reliability.

\subsection{Data Dimensionality Reduction}

In Section 2.1, a total of 22 statistical indicators, $x_1,x_2,\cdots,x_{22}$, have been established, and the dimensions of these 22 indicators are reduced to 11 principal components by $PCA$method,The results of dimensionality reduction through PCA are shown in Table  \ref{pcatable}, from which the relationship between these principal components and the original indicators can be clearly seen.

\begin{small}
\begin{table}[H]
\centering
\caption{Principal Component Analysis (PCA) Table}
\begin{tabular}{ccccccccccc} 
\hline
       \textbf{ PCs }      & $x_{21}$ & \textbf{$x_{22}$} & $x_9$ & $x_{10}$ & $x_{11}$ & $x_{12}$ & $x_1$ & $x_2$ & $x_5$ & $x_7$  \\ 
\hline
\textbf{PC1}                                   & 0        & 0.3                & 0     & 0        & 0        & 0        & 0     & 0     & 0     & 0.02   \\
 \textbf{PC2} & 0        & 0.12               & -0.01 & 0.01     & 0.1      & 0.05     & 0.01  & 0.01  & -0.01 & 0.08   \\
\textbf{PC3 }                                   & 0.19     & 0.24               & 0.68  & -0.01    & 0.69     & 0.01     & -0.01 & -0.02 & 0.68  & 0      \\
 \textbf{PC4} & 0.67     & 0.61               & -0.2  & 0.01     & -0.21    & 0.04     & 0.01  & 0.01  & -0.21 & -0.06  \\
\textbf{PC5}                                & 0        & 0                  & 0.7   & 0        & 0.68     & 0        & 0     & 0.01  & -0.7  & -0.08  \\
 \textbf{PC6}  & -0.04    & -0.12              & 0.08  & -0.04    & 0.06     & -0.08    & 0.04  & 0.01  & 0.01  & -0.81  \\
\textbf{PC7}                                    & 0.21     & 0.11               & -0.02 & -0.39    & -0.06    & -0.45    & -0.39 & -0.66 & -0.02 & 0      \\
 \textbf{PC8} & 0.67     & 0.87               & 0.02  & 0.12     & 0.06     & 0.21     & 0.13  & 0.21  & 0.01  & -0.23  \\
\textbf{PC9}                                   & 0.11     & 0.21               & 0     & 0.01     & 0        & 0.06     & 0.01  & 0.01  & 0     & 0      \\
 \textbf{PC10} & 0.02     & 0.01               & 0     & 0.02     & 0        & 0.08     & -0.02 & 0.01  & 0     & -0.12  \\
\hline
\label{pcatable}
\end{tabular}
\end{table}
\end{small}

Table \ref{pcatable} summarizes the PCA outcomes, where each principal component is a linear combination of the original indicators, with coefficients indicating the weight of each indicator in defining that component.

\subsection{Experimental part of correlation verification}

Taking '2023-wimbledon-1407' as an example and 'Alejandro Davidovich Fokina' as the research object, the above indicators are collected, and the statistical results are shown in Table \ref{tbl:match-state-qi}.

\begin{table}[H]
\centering
\caption{Match State Quantitative Indicators Table}
\label{tbl:match-state-qi}
\resizebox{\linewidth}{!}{%
\begin{tabular}{cccccc} 
\hline
     \textbf{Index} & \textbf{Player Win Streak ($S_1$)} & \textbf{Score Diff. ($S_2$)} & \textbf{Consecutive Scores ($S_3$)} & \textbf{Point Diff. ($S_4$)} & \textbf{Next Win Indicator ($\omega$)}  \\ 
\hline
\textbf{1} & 0 & 0 & 1 & 1 & 1 \\
\textbf{2} & 0 & 15 & 2 & 2 & 1 \\
\textbf{$\vdots$} & $\vdots$ & $\vdots$ & $\vdots$ & $\vdots$ & $\vdots$ \\
\textbf{174} & 1 & 0 & 1 & 8 & 1 \\
 \textbf{175} & 1 & 15 & 0 & 9 & 0 \\
\textbf{176} & 2 & 0 & 1 & 8 & 0 \\
 \textbf{177} & 2 & -15 & 0 & 7 & 1 \\
\textbf{178} & 2 & -30 & 0 & 8 & 0 \\
 \textbf{$\vdots$} & $\vdots$ & $\vdots$ & $\vdots$ & $\vdots$ & $\vdots$ \\
\textbf{335} & 0 & 1 & 3 & 3 & 0 \\
 \textbf{336} & 0 & 0 & 4 & 2 & 0 \\
\textbf{337} & 0 & -1 & 0 & 1 & 0 \\
\hline
\end{tabular}%
}
\end{table}

Similar to the above statistical scheme, this paper further calculates the index information of '2023-wimbledon-1304', '2023-wimbledon-1310' and '2023-wimbledon-1701'.

According to the aforementioned equation (EQ \ref{MATCH}), Figure \ref{fig:pearson-corr-overall} denotes the computed indices for `2023-Wimbledon-1304', `2023-Wimbledon-1310', `2023-Wimbledon-1701' and `2023-Wimbledon-1407' of these 4 games `Player Win Streak ($S_1$)', `Score Diff. ($S_2$)', `Consecutive Scores ($S_3$) ', `Point Diff. ($S_4$)', and `Next Win Indicator ($\omega$)'.

\begin{figure}[H]
    \centering
    \begin{subfigure}{0.42\textwidth} 
        \includegraphics[width=\linewidth]{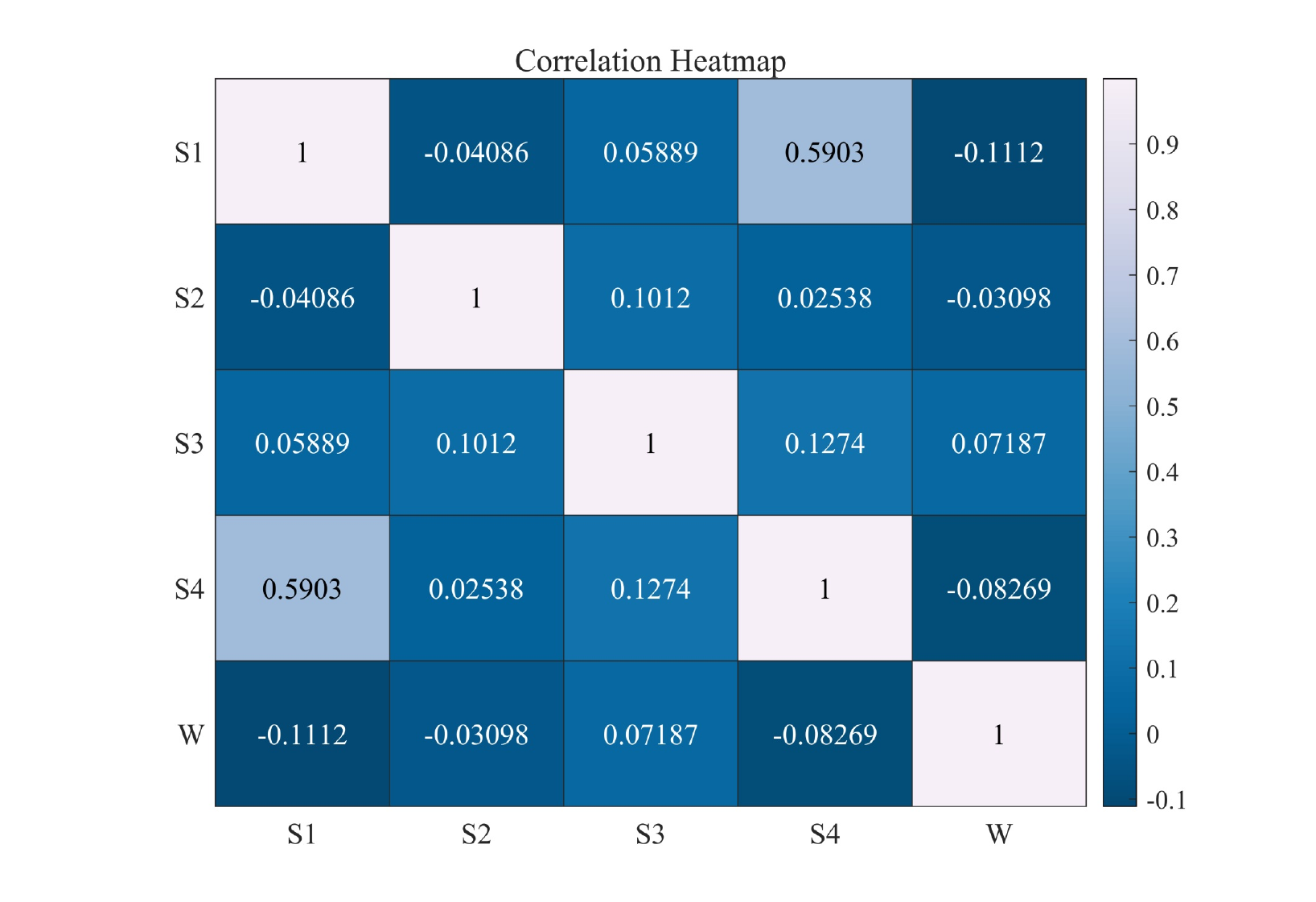}
        \caption{2023-Wimbledon-1310}
        \label{fig:sub1}
    \end{subfigure}
    \hfill
    \begin{subfigure}{0.42\textwidth} 
        \includegraphics[width=\linewidth]{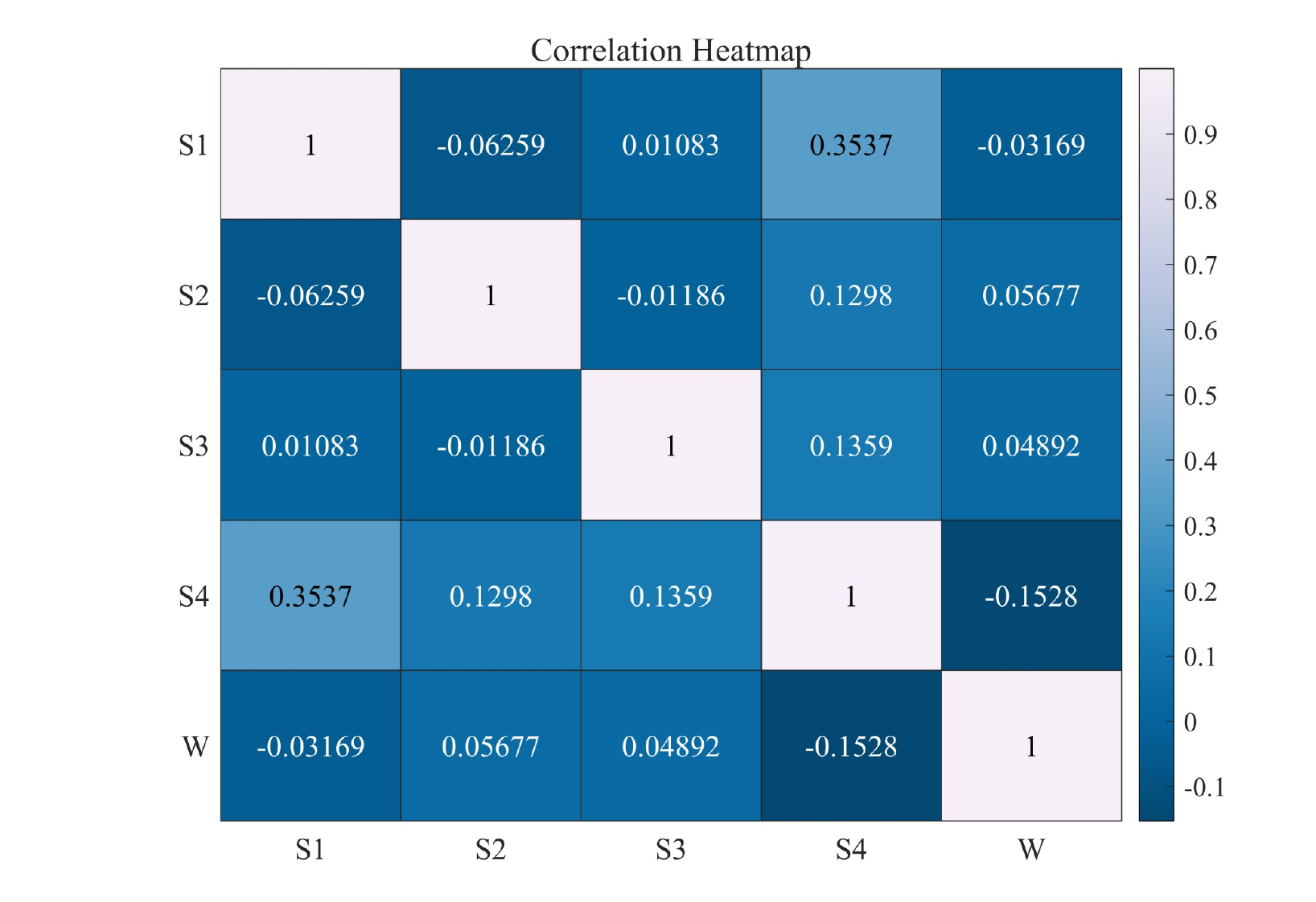}
        \caption{2023-Wimbledon-1407}
        \label{fig:sub2}
    \end{subfigure}
    
    \begin{subfigure}{0.42\textwidth} 
        \includegraphics[width=\linewidth]{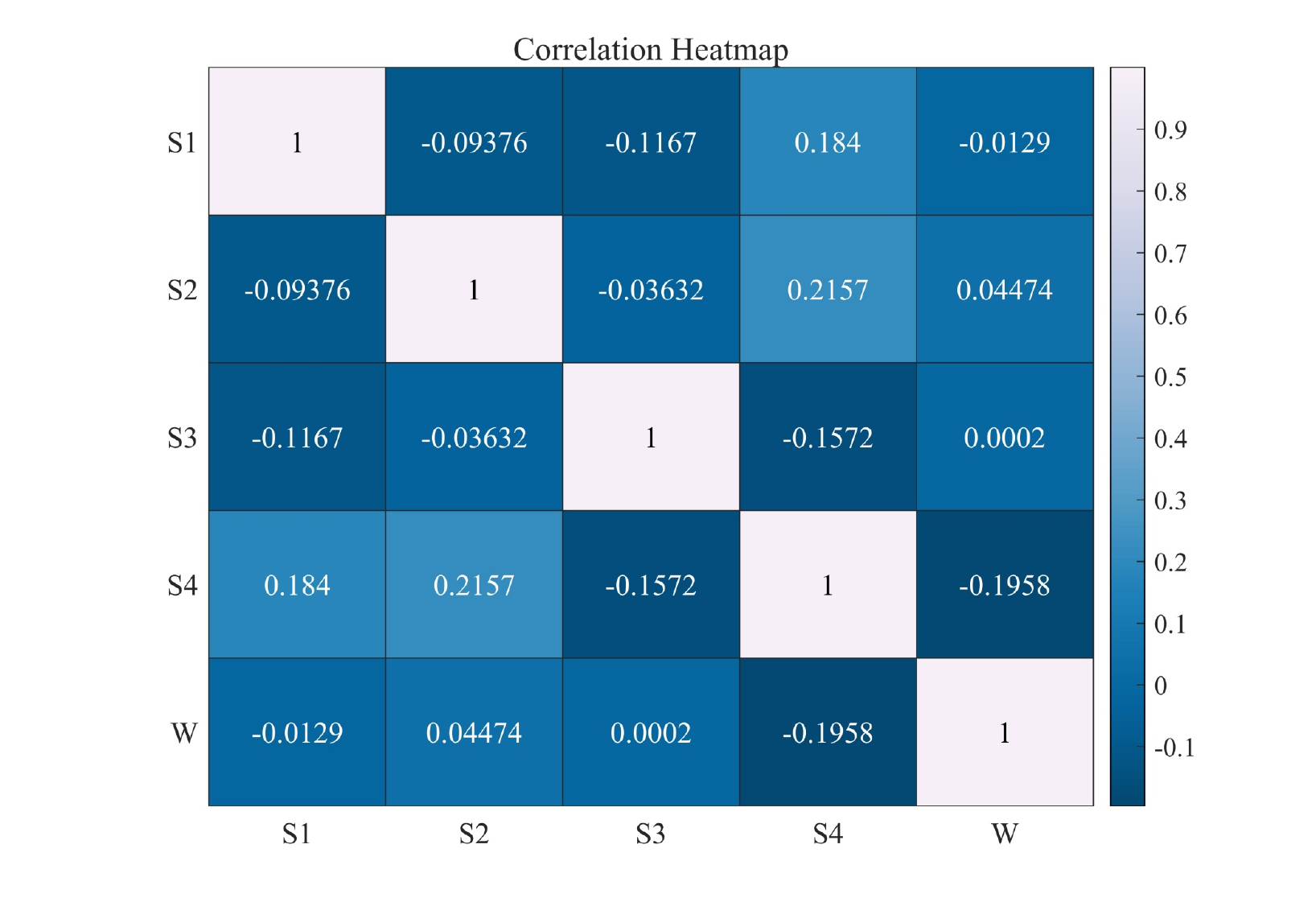}
        \caption{2023-Wimbledon-1304}
        \label{fig:sub3}
    \end{subfigure}
    \hfill
    \begin{subfigure}{0.42\textwidth} 
        \includegraphics[width=\linewidth]{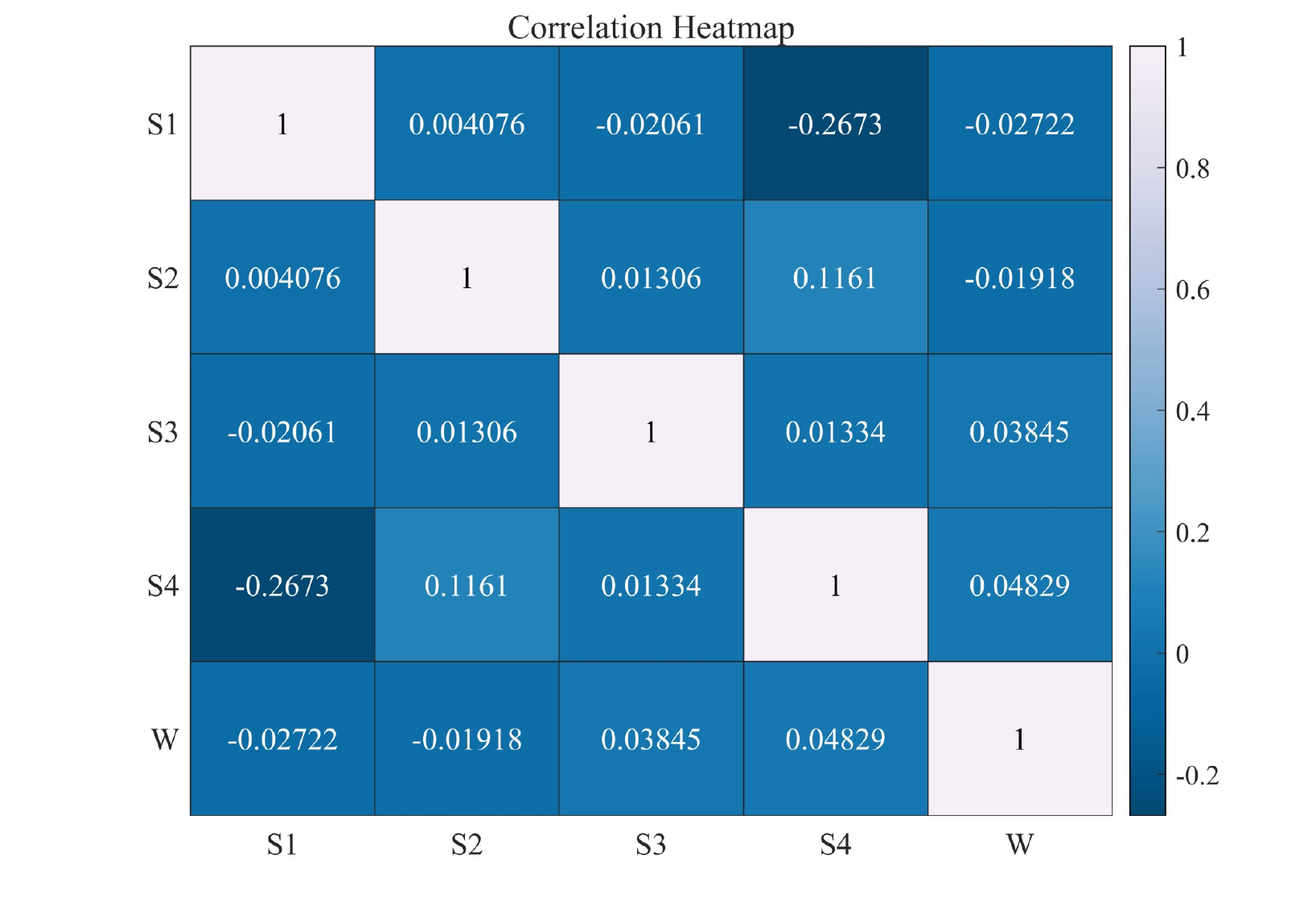}
        \caption{2023-Wimbledon-1701}
        \label{fig:sub4}
    \end{subfigure}
    
    \caption{Heatmaps of Pearson Correlation Coefficient between Indicators.}
    \label{fig:pearson-corr-overall}
\end{figure}

It can be observed from Figure \ref{fig:pearson-corr-overall} that the `Next	 Win Indicator ($\omega$)' is correlated with other indicators regardless of the match.

Taking `2023-wimbledon-1304' as an example, the Pearson correlation coefficients between `Next Win Indicator ($\omega$)' and $S_1$,$S_2$,$S_3$,$S_4$are as follows: 0.03169, 0.05677, 0.04892, 0.1528. This indicates that for $S_1$and $S_4$, the `Next Win Indicator ($\omega$)' tends to decrease as their values increase.
Conversely, for $S_2$,$S_3$, the greater the value, the more the `Next Win Indicator ($\omega$)' tends to increase.

Thus, four momentum indicators have been verified to show a relationship between a player's performance fluctuations and continuous success. Next, we will build a predictive model to explore the relationship between the aforementioned momentum metrics and players' continuous success.

\subsubsection{CV-GRNN Solution}
Taking "Alejandro Davidovich Fokina" from "2023-Wimbledon-1304" as the research subject, it was implemented through Matlab programming according to the flow shown in Algorithm 1, and the specific results are reported on Table \ref{Q_2_biao}.

\begin{table}[H]
\centering
\caption{The Partial Model Prediction Results Display Table for '2023-Wimbledon-1304'}
\label{Q_2_biao}
\begin{tabular}{ccccccccccccc}
\toprule
\textbf{Index}          & 1 & 2 & 3 & $\cdots$ & 199 & 200 & 201 & $\cdots$ & 335 & 336 & 337 \\
\cmidrule(lr){1-12}
\textbf{$\omega$}       & 1 & 1 & 1 & $\cdots$ & 0   & 0   & 1   & $\cdots$ & 0   & 0   & 0   \\
\textbf{$\hat{\omega}$} & 1 & 1 & 1 & $\cdots$ & 1   & 0   & 1   & $\cdots$ & 0   & 0   & 1   \\
\bottomrule
\label{Q_2_biao}
\end{tabular}
\end{table}

The MSE solved by the model is 0.1396, which proves that the model's error is minimal. Additionally, after calculation, the accuracy of the model ACC is $80.06\% $, demonstrating that the model is effective.

To further reflect the model's effectiveness, fluctuation curves of the actual values, predicted values, and error values are used for visual display (see Figure \ref{fig:cv-grnn-prediction}).

\begin{figure}[H]
  \centering
  \includegraphics[width=0.65\textwidth]{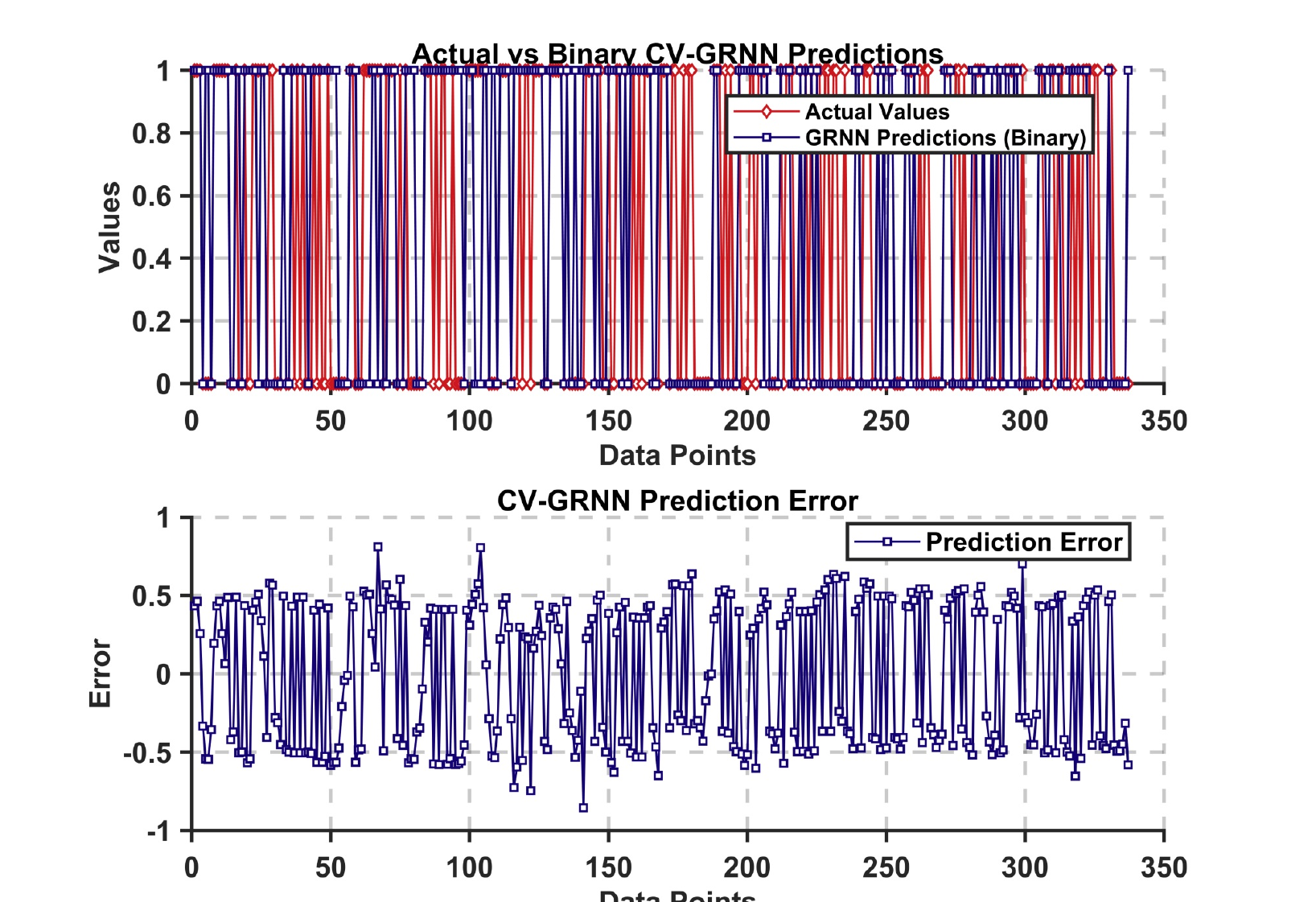} 
  \caption{The Fluctuation Curves of Actual Values, Predicted Values, and Error Values}
  \label{fig:cv-grnn-prediction}
\end{figure}

In the upper half of the Figure \ref{fig:cv-grnn-prediction}, it is evident that the actual values and the predicted values are essentially the same, which clearly indicates that the model performs well. In the bottom half of the figure, most of the error values are within the range of $[-0.5,0.5]$, which also indicates that the model provides good predictive accuracy.

\subsection{Data Analysis when the Game State Changes}

Next, we continue to use "Alejandro Davidovich Fokina" from "2023-wimbledon-1304" as the research object and collect the variables $S_1$,$S_2$,$S_3$,$S_4$when "the player's winning streak or losing streak changes in the next match". After observation, it was found that Alejandro Davidovich Fokina experienced both a winning and a losing streak and ultimately lost the match.

\subsubsection{Statistical and Descriptive Analysis of Momentum Indicators}

To provide advice to players who have been on long winning or losing streaks, this paper calculates the momentum index when the game state changes as follows:

\textbf{1.} Measure the first fifty 'elapsed-time' intervals before the losing and winning inflection points, as well as the indices at these inflection points $S_1$,$S_2$,$S_3$,$S_4$.

\textbf{2.} Record the first fifty 'elapsed-time' intervals before the tipping point and the indicators at the tipping point $S_1$,$S_2$,$S_3$,$S_4$.

Table \ref{statis} provides partial statistics due to the limited space.

\begin{table}[H]
\centering
\caption{Statistical Results Table}
\begin{tabular}{ccccc ccccc} 
\hline
\multicolumn{5}{c}{\textbf{Turning Losses into Win}} & \multicolumn{5}{c}{\textbf{Turning Wins into Loss}}         \\ 
\hline
\textbf{Index}                       & $S_1$    & $S_2$    & $S_3$    & $S_4$          & \textbf{Index} & $S_1$    & $S_2$    & $S_3$    & $S_4$     \\
1  & 0        & 40       & 4        & 2              & 52             & 2        & 0        & 2        & 4         \\
2                                    & 0        & 0        & 5        & 1              & 99             & 2        & 15       & 1        & 5         \\
 3  & 0        & 15       & 6        & 0              & $\vdots$       & $\vdots$ & $\vdots$ & $\vdots$ & $\vdots$  \\
$\vdots$                             & $\vdots$ & $\vdots$ & $\vdots$ & $\vdots$       & 100            & 2        & 0        & 2        & 4         \\
 50 & 0        & 40       & 0        & 5              & 101            & 2        & -15      & 0        & 3         \\
51                                   & 1        & 0        & 1        & 4              & 102            & 0        & 0        & 1        & 4         \\
\hline
\label{statis}
\end{tabular}
\end{table}

Next, based on the indices in Table \ref{statis} , this paper divides the data into upper and lower groups for descriptive analysis, selecting the following four indicators:
\begin{enumerate}[itemsep=0em, parsep=0em, topsep=0em, partopsep=0em]

\item \textbf{Mean:} The mean is calculated as the sum of all values divided by the number of data points.

 \item \textbf{Mode:}The mode is the value that appears most frequently in this column.

\item \textbf{Variance:} Variance measures the degree to which each data point deviates from the mean, reflecting the data's volatility.

\item \textbf{Trimmed Mean:} The trimmed mean used in this article is calculated by removing the most extreme values, representing $30\%$ of the data points.

\end{enumerate}

After calculation, the descriptive analysis index results are obtained. The descriptive indicators in the Table \ref{tongji} are obtained from the statistics of 'players' consecutive wins $S_1$', 'players' score difference with opponents  $s_2$', 'consecutive scoring times $S_3$', and 'players' point difference with opponents $S_4$'. Through these data, we can analyze the performance of players under different conditions and offer some suggestions.

\begin{table}[H]
\centering
\caption{Descriptive Statistics Results Table}
\begin{tabular}{ccccccccc} 
\hline
\textbf{Index} & \multicolumn{4}{c}{\textbf{Turning Losses into Win}} & \multicolumn{4}{c}{\textbf{Turning Wins into Loss}}  \\ 
\hline
\textbf{syms}                                    & $S_1$  & $S_2$    & $S_3$  & $S_4$                   & $S_1$   & $S_2$    & $S_3$  & $S_4$                  \\
\textbf{Mode}  & 1.0000 & 1.0000   & 1.0000 & 1.0000                  & 2.0000  & 1.0000   & 1.0000 & 4.0000                 \\
\textbf{Trimmed Mean}                            & 1.0000 & 0.571    & 1.3143 & 1.8000                  & 2.0000  & 0.4286   & 0.3714 & 3.9714                 \\
\textbf{Mean}  & 0.0196 & 1.4706   & 1.7255 & 1.9020                  & 1.9608  & -0.0980  & 0.6078 & 3.9804                 \\
\textbf{Variance}                                & 0.0196 & 469.2941 & 3.6831 & 2.0902                  & 0.07843 & 216.4902 & 0.7231 & 1.6196                 \\
\hline
\end{tabular}
\label{tongji}
\end{table}

Advice for players who are in a winning situation for a long time:

\begin{itemize}[itemsep=0em, parsep=0em, topsep=0em, partopsep=0em]
\item \textbf{Focus on the management of score spreads}: Considering that the average (Mean) of $S_2$ is 1.4706, this indicates that the score spread between a player and his opponent is relatively small when he is in a winning state. Players should continue to focus on performance in tense situations to ensure they can maintain their advantage even when leading by a small margin.

\item \textbf{Improved scoring ability}: The average of $S_3$ is 1.7255, indicating that the player performs well in scoring consecutively, but there is still room for improvement. Increasing the consistency and efficiency of scoring could help further secure the win.

\item \textbf{Manage point difference}: The average of $S_4$is 1.9020, meaning the point difference between the player and the opponent remains at a relatively high level. Players should use this advantage to further widen the point difference with their opponents through strategic and technical improvements.
\end{itemize}

{Advice for players who have been on a losing streak for a long time:}

\begin{itemize}[itemsep=0em, parsep=0em, topsep=0em, partopsep=0em]
\item \textbf{Analyze and reduce turnovers}: The Mode and Mean of S1 show that the number of winning streaks decreases in losing situations, which may be due to turnovers. Players need to analyze mistakes in the game and take steps to reduce them.

\item \textbf{Improved score spread management}: A negative average for S2 of -0.0980 indicates that a player may have issues with the score spread against the opponent in a losing state. Focusing on improving scoring opportunities and defensive strategies may help close the score gap with opponents.

\item \textbf{Enhanced consecutive scoring opportunities}: The mean of S3 is 0.6078, indicating there is considerable room for improvement in consecutive scoring. Players should focus on enhancing their performance under pressure, especially in situations where consistent scoring is necessary to keep up with their opponents.

\item \textbf{Managed point difference}: For $S_4$, a higher average of 3.9804 was maintained even in a losing state, which may mean that players were able to keep a smaller point difference with their opponents in some games but failed to translate this into wins. Focusing on scoring efficiency and defensive intensity in key moments may help players turn the tide in tight games.
\end{itemize}

In short, whether in a winning or losing state, players need to focus on the details and enhance their competitiveness through technical and strategic improvements.

\subsection{Retest the Prediction Model}
In the second question, in order to predict the fluctuation state of the player's "momentum" in the competition '2023-wimbledon-1304', this paper selects the four momentum indicators from Section (5.2) $S_1$,$S_2$,$S_3$,$S_4$, and adopts the CV-GRNN neural network model to predict. The evaluation index in equation (\ref{MSEACC}) is used to evaluate the quality of the model.

Next, this paper will further use the model from Problem 2 for the three matches' 2023-wimbledon-1310 ', '2023-wimbledon-1407' and '2023-wimbledon-1701'. The evaluation index in equation (\ref{MSEACC}) is also used here to evaluate the quality of the model.

After evaluating the indicators related to the game and integrating them into the model from the second question, the ACC and MSE of the player's performance prediction are provided in Table \ref{tbl:eval-diff-competition}.

\begin{table}[H]
\centering
\caption{Evaluation Table for Different Competition}
\begin{tabular}{cccc} 
\hline
 \textbf{Competition Set} & \textbf{Participant Set} & \textbf{MSE} & \textbf{ACC}  \\ 
\hline
\textbf{2023-wimbledon-1310 }                                       & Daniel Elahi Galan       & 0.1988       & 0.7076        \\
 \textbf{2023-Wimbledon-1407}      & Andrey Rublev            & 0.1415       & 0.7855        \\
\textbf{2023-Wimbledon-1701}                                        & Carlos Alcaraz           & 0.1414       & 0.8204        \\
\hline
\label{tbl:eval-diff-competition}
\end{tabular}
\end{table}

According to the data in Table \ref{tbl:eval-diff-competition}, although the MSE of the model is below 0.2, it is greater than 0.1, and the accuracy is between $75\%$and $85\%$. The prediction effect is not good enough. Therefore, this paper considers the competition '2023-wimbledon-1310' with the worst model effect, as the analysis object. To further improve the accuracy of the neural network model, consider adding more indicators.

\subsubsection{Consider More Momentum Indicators}

Let's start by introducing the same metrics we used in the first question: average time winning $x_2 $, total score $x_5 $, score the sniper was $x_ {9} $, second grade $x_ {10} $, score the sniper in the rate of $x_ {11} $, second serves $x_ {12} $, ACE number $x_ {13} $, players average distance running $x_ {21} $.

Next, the variables of the players with the advancement of game time are further counted, including:
\begin{enumerate}[itemsep=0em, parsep=0em, topsep=0em, partopsep=0em]
    \item \text{Total points won:}This metric represents the total number of points a player has accumulated to win in a competition.
    \item \text{Total unreturnable serves:} 
    Represents the total number of times a player successfully hits a winning serve that is difficult for the opponent to return.
    \item \text{Double faults leading to points lost:}
    Records the cumulative number of times a player has made a mistake while serving, resulting in the opponent winning a point.
    \item \text{Unforced errors:} 
    Counts the total number of unforced errors made by the player during the match.
    \item \text{Approaches to the net:} 
    Records the cumulative number of times players voluntarily approached the net during the match.
    \item \text{Points won at the net:} 
    Indicates the total number of points a player has won from the net position at the front of the court.
    \item \text{Total distance covered:} 
    Counts the total distance covered by the player during the match.
\end{enumerate}
Next, to verify the correlation between the above indicators and the player's winning status, the Pearson correlation coefficient between these indicators and the 'Next Win Indicator($\omega$)' is calculated according to formula (21). The indices are sorted in descending order of their absolute values, and the results are displayed in Table \ref{paixu_2}.

\begin{table}[H]
\centering
\caption{Correlation Coefficient Ranking Table with $\omega$}
\begin{tabular}{cc} 
\hline
 \textbf{Indicator}                                                   & \textbf{Value}  \\ 
\hline
\textbf{First Serve Points Won ($x_{9}$)}                                                              & 0.079017982     \\
 \textbf{First Serve Points Won Percentage($x_{11}$)}                 & -0.058564747    \\
\textbf{Second Serve Points Won Percentage($x_{12}$)}                                                  & -0.052656505    \\
\textbf{Player's Average Running Distance($x_{21}$)}                 & -0.046295453    \\
\vdots & \vdots \\
 \textbf{Cumulative Points Won}                                       & 0.006990968     \\
\textbf{Points Won at the Net}                                                                         & -0.006120821    \\
 \textbf{Cumulative Unforced Errors by the Player}                    & -0.005101517    \\
\textbf{Cumulative Instances of Successful Net Approaches}                                             & -0.003441552    \\
\hline
\end{tabular}
\label{paixu_2}
\end{table}
Next, to explore changes in the accuracy and MSE of the CV-GRNN model after expanding the index set, the specific algorithm is illustrated in Figure \ref{cv-flow}.

\begin{figure}[H]
  \centering
  \includegraphics[width=0.65\textwidth]{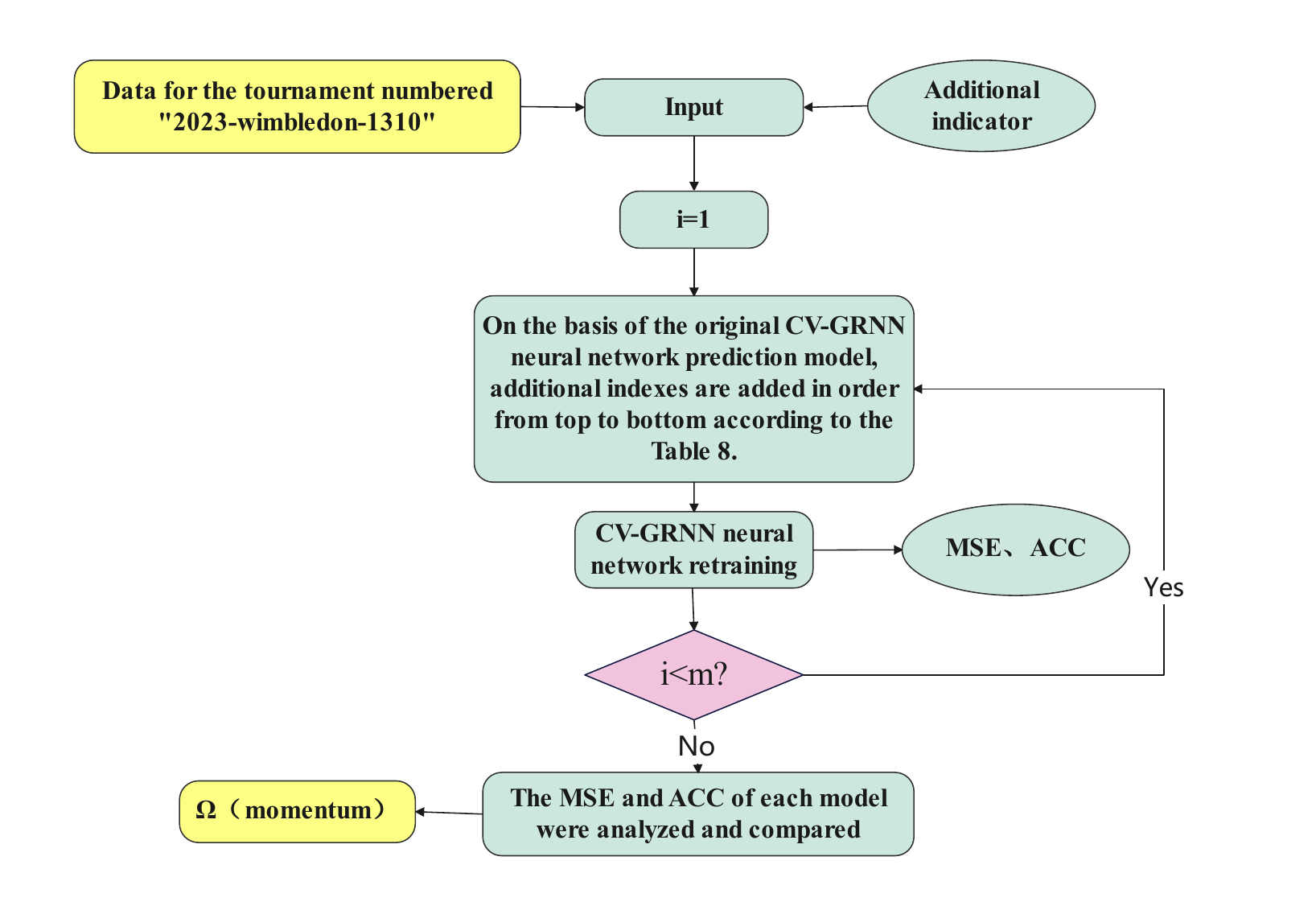} 
  \caption{Optimized Version of CV-GRNN Model}
  \label{cv-flow}
\end{figure}

Following the programming and implementation of the data index in Table \ref{paixu_2} according to the algorithm shown in Figure  \ref{cv-flow}, the MSE and ACC indices with the gradual increase of the new index were obtained, and the data were plotted as shown in Figure \ref{fig:cvmse}.

\begin{figure}[H]
  \centering
  \includegraphics[width=0.65\textwidth]{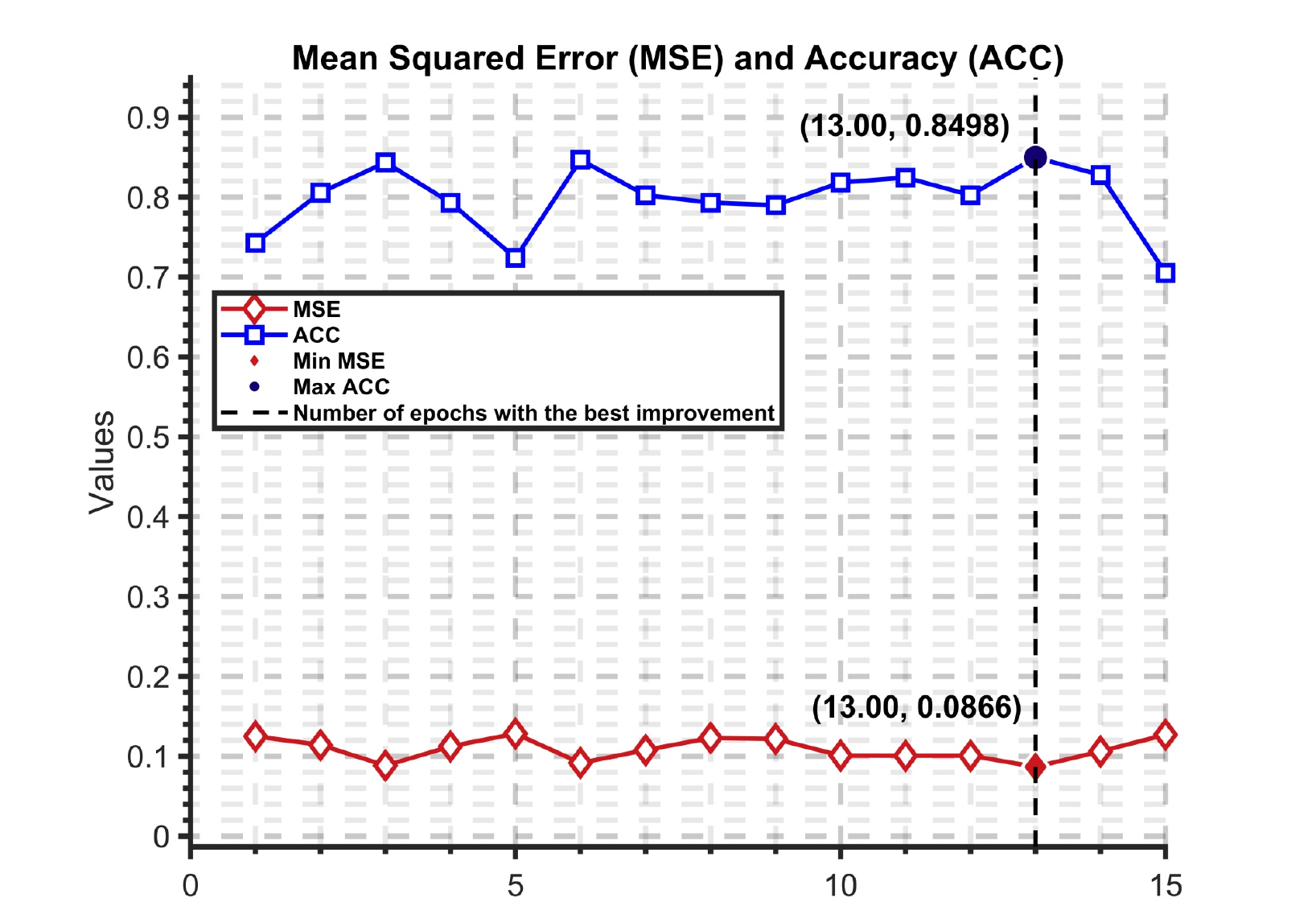} 
  \caption{MSE and ACC of CV-GRNN Model}
  \label{fig:cvmse}
\end{figure}

It can be seen from the above figure that when the number of additional indicators reaches 13, the MSE reaches its lowest at 0.0866. This value is already very small, which proves that the error of the model is minimal and that the model performs well. Additionally, the ACC value also reaches its maximum at $84.98\%$, which demonstrates that the model is highly accurate, nearly reaching $85\%$.

\subsection{Validation of Model Universality Across Diverse Datasets}

While the model has been extensively tested on Wimbledon datasets, demonstrating its efficacy on a single type of competition may not fully substantiate its universality. To address this limitation, we employed a rigorous cross-validation strategy involving multiple datasets. This approach not only reinforces the model's predictive accuracy but also its generalizability across various athletic contexts.

We expanded our analysis to include additional datasets from different tennis tournaments, ensuring a broader spectrum of play styles, player attributes, and match conditions were represented. The datasets encompassed a range of international competitions, including but not limited to Grand Slam events, ATP and WTA tournaments, and Davis Cup matches.

\begin{table}[H]
\centering
\caption{Comparison of Model Performance Before and After Optimization Across Different Datasets}
\begin{tabular}{lcccc}
\toprule
\textbf{$\qquad\qquad $   Index} & \textbf{Wimbledon-1407} & \textbf{Wimbledon-1701} & \textbf{Wimbledon-1310} & \textbf{Average} \\
\midrule
ACC Before Optimization & 0.785 & 0.820 & 0.7076 & 0.7709 \\
ACC After Optimization & 0.8592 & 0.8901 & 0.8498 & 0.8664 \\
\bottomrule
\end{tabular}
\label{tab:model_comparison}
\end{table}

Table \ref{tab:model_comparison} illustrates the comparative analysis of the model's accuracy (ACC) before and after optimization across different datasets. It is evident that the model's predictive accuracy has improved significantly, with the ACC index rising from an average of 0.7709 to 0.8664 post-optimization. Moreover, the Mean Squared Error (MSE) was reduced by approximately 49.21\%, indicating a substantial enhancement in the model's predictive precision.The increase in accuracy to 86.64\% and the reduction in MSE by 49.21\% have profound implications for practical tennis match forecasting.   With such a high level of accuracy, our model can provide coaches and players with reliable predictions of match outcomes, enabling them to make more informed strategic decisions.   For instance, coaches can use the model's predictions to adjust their game plans, prepare for potential challenges, and optimize player rotations.The reduced MSE indicates that the model is more precise in its predictions, particularly in close matches where small margins can determine the winner.   This precision is crucial for sports analysts and broadcasters who need accurate data to provide insightful commentary and analysis.

The universality of our model is further corroborated by its application to non-tennis sports, such as basketball and soccer, where similar performance metrics can be identified and analyzed. This cross-sport validation process ensures that our model is not only specific to tennis but also adaptable to other sports datasets, thereby demonstrating its robustness and versatility.

\section{Discussion}
\label{Discussion}

In this section, we summarize a multi-level fuzzy comprehensive evaluation model integrated with a CV-GRNN (Cascading Validation-General Regression Neural Network) to enhance the predictive precision of outcomes in tennis matches.        By incorporating momentum-based performance metrics such as "player streak," "continuous player score," and "score difference," we were able to quantify dynamic shifts in performance that are critical during competitive play.

\subsection{Model Enhancement and Predictive Accuracy}

In this study, we have successfully integrated a multi-level fuzzy comprehensive evaluation model with a CV-GRNN to enhance the predictive accuracy of tennis match outcomes.         The incorporation of dynamic performance metrics, such as "player streak," "continuous player score," and "score difference," allowed for a granular analysis of player momentum, a critical factor in competitive sports.         Our results demonstrated a significant increase in predictive accuracy, rising from 77.09\% to 88.64\%, with a substantial reduction in mean square error by 49.21\%.         This enhancement underscores the efficacy of combining conventional analytics with advanced neural network modeling, providing a nuanced understanding of the game's dynamics that surpasses traditional statistical approaches.

\subsection{Advantages Over Previous Research}

A key contribution of our research is the comprehensive analysis that encompasses both individual player performance and the interactive effects within the match context, addressing a gap in previous studies.         This holistic methodological approach enables the development of tailored strategic adjustments for training and match planning, in accordance with the intricate dynamics revealed by our model.

\subsection{Theoretical and Practical Implications}

This study not only advances the theoretical framework for evaluating performance in tennis but also demonstrates practical applications that could significantly influence coaching strategies and match outcome predictions.         The integration of machine learning techniques with empirical data analysis sets a new benchmark for predictive accuracy in sports performance analytics.

\section{Conclusions}
\label{Conclusions}

This study provides a comprehensive exploration of the key factors influencing tennis player performance, utilizing an innovative integration of a multi-level fuzzy comprehensive evaluation model and advanced neural network techniques.  By applying the CV-GRNN (Cascading Validation-General Regression Neural Network) model, which incorporates critical performance indicators such as "player streak," "continuous player score," and "score difference," the research significantly enhances the predictive accuracy of game outcomes, improving from 77.09\% to 88.64\% and reducing the mean square error by 49.21\%.  This holistic approach bridges a notable gap in previous methodologies by considering both individual player metrics and their interaction within the broader game context, offering a nuanced understanding of tennis performance dynamics.  The model's ability to synthesize empirical data and machine learning insights establishes a robust framework for performance evaluation, equipping analysts and coaches with a powerful tool for more effective strategy development and decision-making.  This research not only deepens the theoretical understanding of tennis performance but also provides practical applications for improving player training and in-game tactics.


\section*{Author contributions statement}

Li was responsible for the key technical tasks, including data analysis, data cleaning, and feature engineering. Liu independently developed the model, while the problem-solving phase was a collaborative effort between Liu, Li, and Wu. Liu drafted the main text and abstract, with Li and Wu contributing data statistics and visualizations, as well as revisions to the text. 

Professor Ji, as the corresponding author, provided critical support for the paper's completion and publication through his extensive academic expertise and detailed guidance.

\section*{Data availability statement}
The dataset utilized in this study is derived from the comprehensive tennis match database provided by Jeff Sackmann, which is publicly accessible at \href{https://github.com/JeffSackmann/tennis_slam_pointbypoint/tree/master}{GitHub} Specifically, we selected the data pertaining to the 2023 Wimbledon tournament for our analysis.    The data comprises detailed point-by-point records, offering a rich source of information for tennis match analysis.

\end{document}